\documentclass[conference]{IEEEtran}
\PassOptionsToPackage{protrusion, expansion}{microtype} 

\usepackage{amsmath, amsthm}
    
\usepackage{graphicx, xcolor} 
\usepackage{url}
\usepackage{breakurl}
\usepackage[breaklinks]{hyperref}
\usepackage{adjustbox}
\usepackage{booktabs}
\usepackage{multirow}
\usepackage{caption}
\usepackage{subcaption}
\usepackage{enumitem}
\usepackage{lipsum}
\usepackage{setspace}
\usepackage{booktabs}
\usepackage{multirow}
\usepackage{algorithm, algpseudocode}
    \definecolor{BLUE}{rgb}{.0, .2, .6}
    \definecolor{BLUEalt}{HTML}{1e50a2}
    \definecolor{RED}{HTML}{c9171e}
    \algrenewcommand{\alglinenumber}[1]{{\scriptsize\bfseries\ttfamily\color{RED}#1}}

\usepackage{footnote}
\makesavenoteenv{tabular}
\makesavenoteenv{table}
\usepackage{cite}
\IEEEoverridecommandlockouts

\usepackage[acronym]{glossaries}
\makeglossaries
 
\newglossaryentry{fclayer}{
    name={\texttt{fc}-layer},
    description={fully connected layer}
}
 
\newglossaryentry{szfull}{
    name=\textsc{SZ lossy compression},
    description={SZ lossy compression}
}

\newglossaryentry{deepsz}{
    name=\textsc{DeepSZ},
    description={DeepSZ}
}

\newglossaryentry{dataArray}{
    name=\textsc{data array},
    description={}
}

\newglossaryentry{indexArray}{
    name=\textsc{index array},
    description={}
}
 
\newacronym{gcd}{GCD}{Greatest Common Divisor}
\newacronym{lcm}{LCM}{Least Common Multiple}

\begin{document}
%

\title{Understanding GPU-Based Lossy Compression for Extreme-Scale Cosmological Simulations}

\author{\IEEEauthorblockN{Sian Jin,\IEEEauthorrefmark{1}
Pascal Grosset,\IEEEauthorrefmark{2}
Christopher~M. Biwer,\IEEEauthorrefmark{2}
Jesus Pulido,\IEEEauthorrefmark{2} 
Jiannan Tian,\IEEEauthorrefmark{1}
Dingwen Tao,\IEEEauthorrefmark{1}\thanks{Corresponding author: Dingwen Tao, Department of Computer Science, The University of Alabama, Tuscaloosa, AL 35487, USA.}
and James Ahrens\IEEEauthorrefmark{2} 
}
\IEEEauthorblockA{\IEEEauthorrefmark{1}
The University of Alabama, AL, USA\\ \{sjin06, jtian10\}@crimson.ua.edu, tao@cs.ua.edu}
\IEEEauthorblockA{\IEEEauthorrefmark{2}Los Alamos National Laboratory, NM, USA\\
\{pascalgrosset, cmbiwer, pulido, ahrens\}@lanl.gov}}

\maketitle

\begin{abstract}
To help understand our universe better, researchers and scientists currently run extreme-scale cosmology simulations on leadership supercomputers. However, such simulations can generate large amounts of scientific data, which often result in expensive costs in data associated with data movement and storage. 
Lossy compression techniques have become attractive because they significantly reduce data size and can maintain high data fidelity for post-analysis.
In this paper, we propose to use GPU-based lossy compression for extreme-scale cosmological simulations. Our contributions are threefold: (1) we implement multiple GPU-based lossy compressors to our open-source compression benchmark and analysis framework named Foresight; (2) we use Foresight to comprehensively evaluate the practicality of using GPU-based lossy compression on two real-world extreme-scale cosmology simulations, namely HACC and Nyx, based on a series of assessment metrics; and (3) we develop a general optimization guideline on how to determine the best-fit configurations for different lossy compressors and cosmological simulations. Experiments show that GPU-based lossy compression can provide necessary accuracy on post-analysis for cosmological simulations and high compression ratio of 5$\sim$15$\times$ on the tested datasets, as well as much higher compression and decompression throughput than CPU-based compressors.
\end{abstract}

\section{Introduction}
\label{sec:intro}

Today's scientific simulations on leadership supercomputers  play important roles in many science and engineering domains such as cosmology. Cosmological simulations enable researchers and scientists to investigate new fundamental astrophysics ideas, develop and evaluate new cosmological probes, and assist large-scale cosmological surveys and investigate systematic uncertainties~\cite{heitmann2019hacc}. 
Given the scale of such surveys and the demands on high-accuracy predictions, such cosmological simulations are usually computation-intensive and must be run on leadership supercomputers. 
Today's supercomputers have evolved to heterogeneity with accelerator-based architectures, in particular GPU-based high-performance computing (HPC) systems, such as the Summit system~\cite{summit} at Oak Ridge National Laboratory.
To adapt to this evolution, cosmological simulation code such as HACC (the Hardware/Hybrid Accelerated Cosmology Code) and Nyx~\cite{almgren2013nyx} (an adaptive mesh cosmological simulation code) are particularly designed for GPU-based HPC systems and can be efficiently scaled to simulating trillions of particles on millions of cores~\cite{habib2013hacc, habib2016hacc}. 

However, extreme-scale cosmological simulations produce huge amounts of data, and as a result storage size and bandwidth have become major bottlenecks~\cite{wan2017comprehensive,wan2017analysis,cappello2019use}. 
For example, one HACC simulation for trillions of particles can generate up to 220 TB of data for each snapshot. Thus, a total of 22 PB of data would be generated, assuming 100 snapshots during the simulation.
Moreover, the I/O time would exceed 10 hours, taking into consideration a sustained storage bandwidth of 500 GB/s, which is deterred. 
To address these issues, the data are usually saved using a process known as decimation. Decimation stores one snapshot every other time step during the simulation. This process can lead to a loss of valuable simulation information and negatively influence the post-analysis. 
A better solution to this simple decimation strategy has been proposed---a new generation of error-bounded lossy compression techniques such as SZ~\cite{tao2017significantly, di2016fast, liangerror} and ZFP~\cite{zfp}.
These error-bounded lossy compression techniques can usually achieve much higher compression ratios, given the same distortion, as demonstrated in many prior studies~\cite{di2016fast,tao2017significantly,zfp,liangerror,lu2018understanding,luo2019identifying,tao2018optimizing,cappello2019use}.

Although lossy compression can significantly reduce data size, an open question remains on how to guarantee the data distortion caused by the compression is acceptable.
For instance, dark matter halos~\cite{eke2001power} play an important role in the formation and evolution of galaxies and hence cosmological simulations. None of the existing work, however, is dedicated to the study of the alterations of dark matter halos caused by different lossy compressors on cosmological simulation datasets. 
Instead, most of the existing work focuses on general data distortion metrics, such as peak signal-to-noise ratio (PSNR), normalized root-mean-square error, Mean Relative Error (MRE), and Mean Square Error (MSE), all of which cannot satisfy the demands of cosmological simulation developers and users. 
On the other hand, as a result of the evolution of supercomputer architecture (e.g., more powerful GPUs on a single node, high-speed CPU-GPU and direct GPU-GPU interconnects), a recent study of cosmological simulations on the Summit supercomputer \cite{hacc-summit} shows a significant performance improvement compared to prior studies \cite{habib2013hacc,habib2016hacc}. 
Specifically, the HACC simulation with 0.1 trillion particles using 1,024 Summit nodes can take about 10 seconds per timestep and generate 2.5 TB per snapshot, whereas the SZ lossy compressor, for instance, can achieve a throughput of about only 2 TB/s with 1,024 nodes (64 Intel Xeon 6148 CPU cores per node), according to prior work \cite{tao2017significantly, sz-openmp-report} and our experiment.
Even ignoring the time to transfer the uncompressed data from the GPU to the CPU, CPU-based lossy compressors would still cause about more than 10\% overhead of the overall performance, which would limit the I/O performance gain by lossy compression.
Thus, several lossy compressor development teams have recently released the GPU versions to reduce the compression overhead. These GPU versions can both accelerate the compression computation and reduce the time needed to transfer the data from GPU to CPU after the compression.
However, there has been no prior work studying GPU-based lossy compression for large-scale cosmological simulations such as HACC and Nyx. Hence, a comprehensive evaluation and study on such simulations is crucial to the community. 
To the best of our knowledge, this paper is the first work on quantitatively evaluating GPU-based lossy compression on real-world extreme-scale cosmological simulations. Our work takes into consideration cosmology-specific evaluation metrics while developing an evaluation methodology and a framework to assist analysis on complex results designed to provide useful insights.

Our study can benefit three target audiences: 
(1) cosmological simulation developers and users who intend to leverage lossy compressor to reduce the I/O and storage burden;
(2) lossy compressor developers who target cosmological simulations or other similar large-scale scientific simulations; and
(3) HPC system administrators who intend to deploy lossy compressors as a system data-management tool.
Moreover, our work can also be applied to other large-scale scientific simulations, which are facing the challenge to understand the impacts of lossy compression on their domain-specific metrics, such as climate simulation \cite{cesm-simulation} with structural similarity index \cite{baker2014methodology}.
Our contributions are listed as follows.
\begin{itemize}
    \item We carefully implement advanced GPU-based lossy compressors into Foresight, our open-source compression benchmark and analysis framework. 
    \item We comprehensively evaluate the practicality of using leading GPU-based lossy compressors (i.e., GPU-SZ and cuZFP) with various compression configurations on two well-known extreme-scale cosmological simulation (i.e., HACC and Nyx) datasets. 
    \item We develop a general optimization guideline for domain scientists on how to determine the best-fit compression configurations for different GPU-based lossy compressors and cosmological simulations.
\end{itemize}

The rest of the paper is organized as follows.
In Section~\ref{sec:background}, we discuss the background information. 
In Section~\ref{sec:problem}, we formulate the research problem.
In Section~\ref{sec:design}, we describe our evaluation methodologies, including our evaluation framework and configurations.  
In Section~\ref{sec:evaluation}, we present the experimental evaluation results on real-world cosmological simulation datasets and our optimization guideline.
In Section~\ref{sec:conclusion}, we present our conclusion and discuss future work.

\section{Background}
\label{sec:background}

In this section, we present background information about advanced lossy compression for scientific data and cosmological simulations. 

\subsection{GPU-Based Lossy Compression for Scientific Data}

\begin{figure*}[h]
\centering
\vspace{-4mm}
\begin{subfigure}{0.20\linewidth}\centering
    \vspace{-3mm}
    \includegraphics[width=3.4cm]{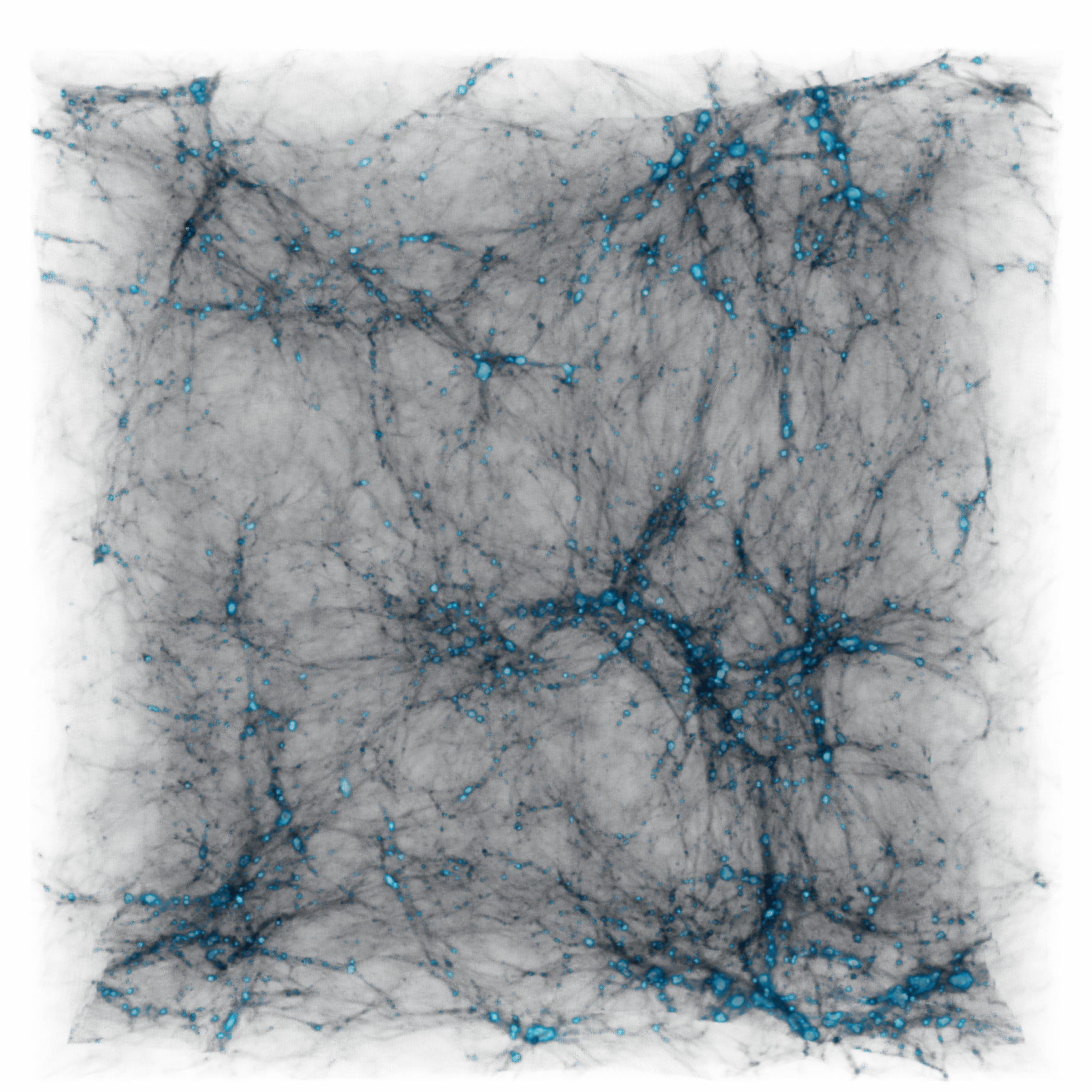}
	\caption{\footnotesize\centering Original}\label{fig:subfigure_orig_crop}
\end{subfigure}
\begin{subfigure}{0.20\linewidth}\centering
    \includegraphics[width=3.4cm]{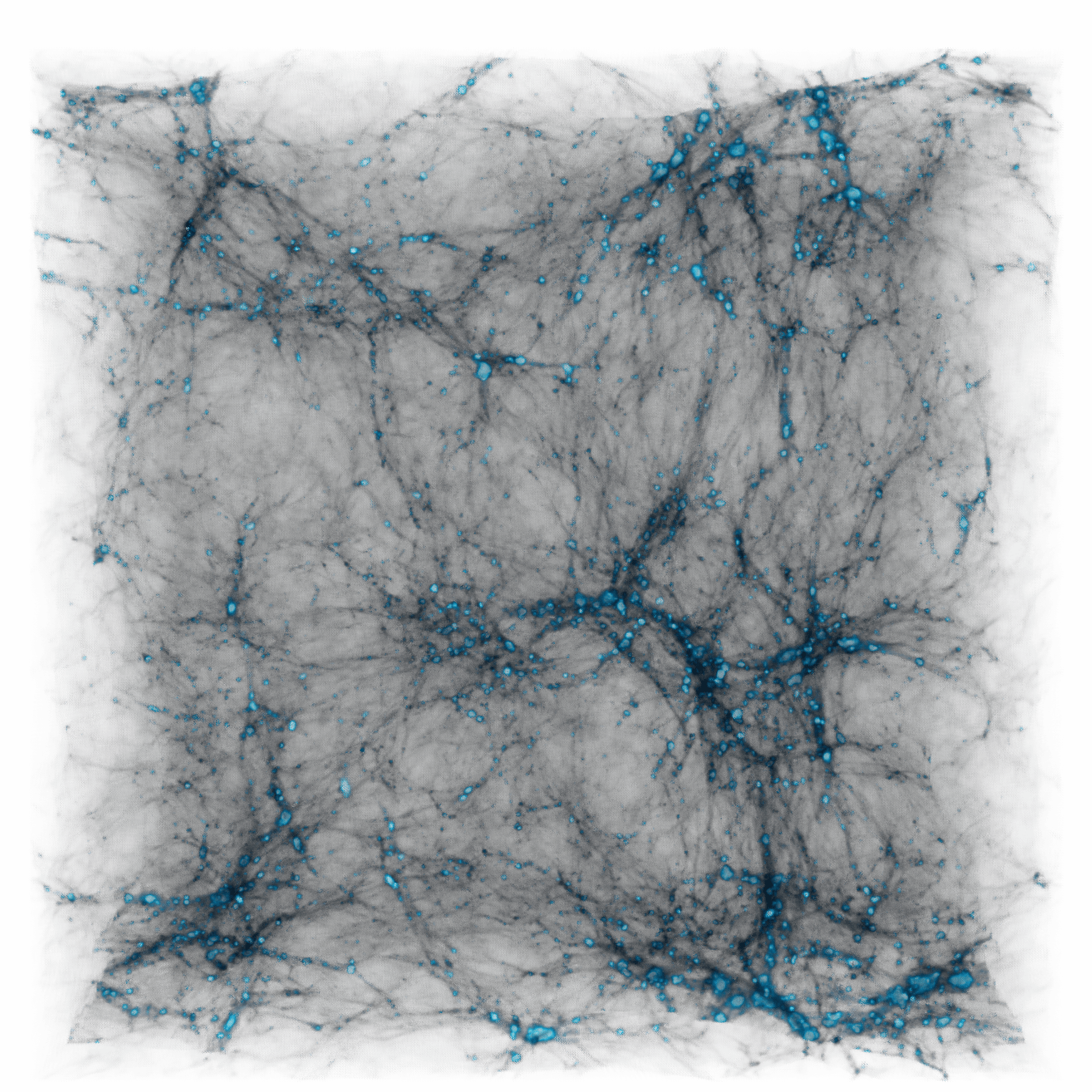}
	\caption{\footnotesize\centering Reconstructed with PW\_REL = 0.1}
	\label{fig:subfigure_sz_rel_error_0.1}
\end{subfigure}
\begin{subfigure}{0.20\linewidth}\centering
    \includegraphics[width=3.4cm]{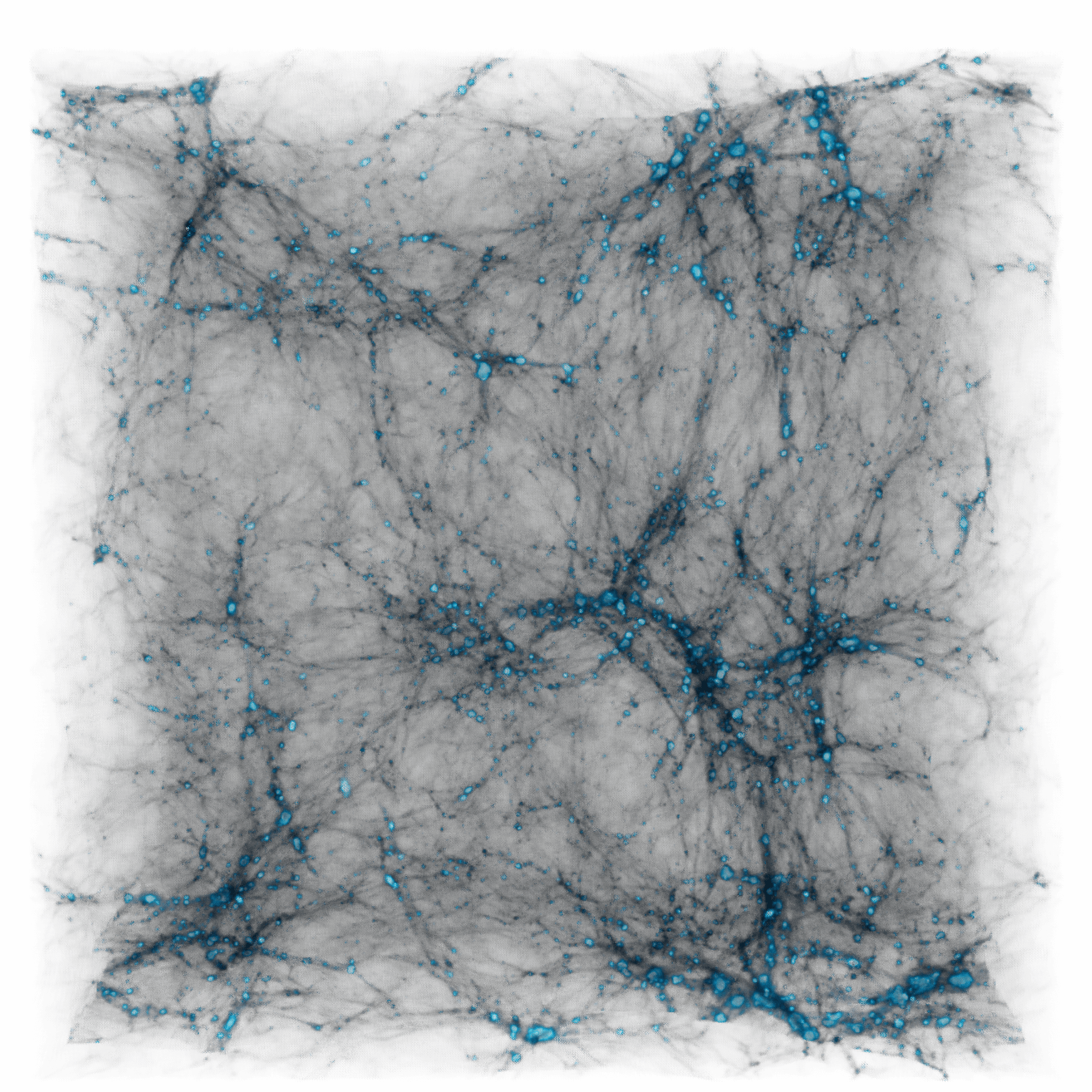}
	\caption{\footnotesize\centering Reconstructed with PW\_REL = 0.25}
	\label{fig:subfigure_sz_rel_error_0.25}
\end{subfigure}
\begin{subfigure}{0.35\linewidth}\centering
    \vspace{-3mm}
    \includegraphics[width=5.8cm]{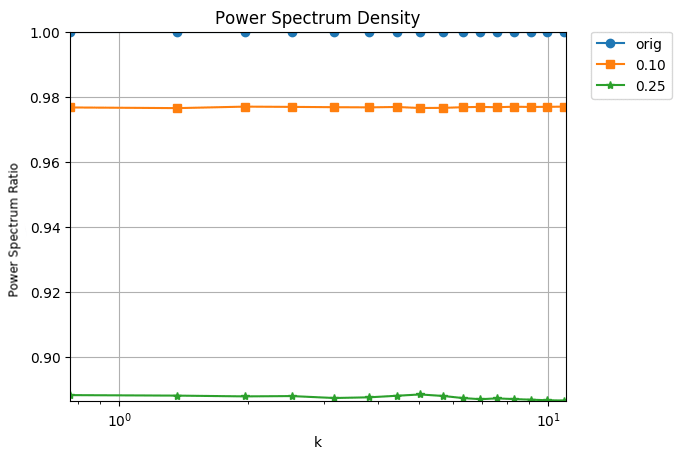}
    \vspace{-3mm}
	\caption{\footnotesize Power Spectrum Density}\label{fig:psd}
\end{subfigure}
\caption{Visualization of (a) original Nyx data, (b) reconstructed Nyx data using GPU-SZ with PW\_REL = 0.1, (c) reconstructed Nyx data using GPU-SZ with PW\_REL = 0.25, and (d) power spectrum density of (a), (b), and (c). }
\label{fig:lossy_compression_sample}
\vspace{-4mm}
\end{figure*}

Scientific data compression has been studied for decades. There are two categories of data compression: lossless and lossy. Lossless compressors such as FPZIP~\cite{lindstrom2006fast} and FPC~\cite{FPC} can provide only compression ratios typically lower than 2:1 for dense scientific data because of the significant randomness of the ending mantissa bits~\cite{son2014data}.

As its name suggests, lossy compression can compress data with little information loss in the reconstructed data. 
Compared to lossless compression, lossy compression can provide a much higher compression ratio while still maintaining useful information in the data. 
Different lossy compressors can provide different compression modes, such as error-bounded mode and fixed-rate mode. 
Error-bounded mode requires users to set an error bound, such as absolute error bound and point-wise relative error bound. The compressor ensures the differences between the original data and the reconstructed data do not exceed the user-set error bound.
Fixed-rate mode means that users can set a target bitrate, and the compressor guarantees the actual bitrate of compressed data to be lower than the user-set value.

In recent years, a new generation of lossy compressors for scientific data has been proposed and developed, such as SZ and ZFP. SZ~\cite{di2016fast, tao2017significantly, liangerror} is a prediction-based error-bounded lossy compressor for scientific data. SZ has three main steps: (1) predict each data point's value based on its neighboring points by using an adaptive, best-fit prediction method; (2) quantize the difference between the real value and predicted value based on the user-set error bound; and (3) apply a customized Huffman coding and lossless compression to achieve a higher compression ratio.
ZFP~\cite{zfp} is another advanced lossy compressor for scientific data that can support both error-bounded mode and fixed-rate mode. 
ZFP first divides the dataset into small blocks (e.g., 4$\times$4) and then compresses each block with the following steps: exponent alignment, orthogonal transform, fixed-point integer conversion, and bit-plan-based embedded coding.

SZ and ZFP were first developed for CPU architectures, and both started rolling out their GPU-based lossy compression recently.
The SZ team has implemented its GPU prototype using OpenMP~\cite{chandra2001parallel}, whereas the ZFP team has released the CUDA implementation of ZFP compression~\cite{cuZFP}.
Compared to lossy compression on CPUs, GPU-based lossy compression can provide much higher throughput for both compression and decompression, as demonstrated in our experimental evaluation addressed in Section~\ref{sec:evaluation}.

\subsection{Cosmological Simulations at Extreme Scale}

HACC and Nyx are two leading cosmological simulations designed to target the extreme-scale supercomputers. According to prior studies~\cite{nyx,habib2016hacc}, HACC and Nyx can run up to millions of cores in the leadership supercomputers in the United States, such as Summit~\cite{summit}. These two simulations complement each other, although each has different area of specialization. 

A high-performance cosmology code HACC simulates the mass evolution of the universe for all available supercomputer architectures~\cite{heitmann2019hacc}.
HACC solves an N-body problem involving domain decomposition, a grid-based medium-/long-range force solver based on a particle-mesh method, and a short-range force solver based on a particle-particle algorithm. The particle-mesh solver is common to all architectures, whereas the short-range solver is architecture-specific. 
For parallelization, HACC uses MPI for the long-range force calculation and architecture-specific programming language for the short-range force algorithms, such as OpenMP and CUDA.

Nyx is an adaptive mesh, hydrodynamics code designed to model astrophysical reacting flows on HPC systems~\cite{almgren2013nyx,nyx}. This code models dark matter as discrete particles moving under the influence of gravity. 
The fluid in gas-dynamics is modeled using a finite-volume methodology on an adaptive set of 3-D Eulerian grids/mesh. 
The mesh structure is used to evolve both the fluid quantities and the particles via a particle-mesh method.
Similar to HACC, Nyx uses MPI and OpenMP for parallelization.

Considering cosmological simulation data format, HACC and Nyx are also two representatives. 
HACC data contains multiple 1-D arrays to represent particles' information, with six of them storing each particle's position $(x, y, z)$ and velocity $(v_x, v_y, v_z)$. In addition to cosmological simulations, this format is commonly used for other N-body simulations, such as molecular dynamics simulation. 
Unlike HACC data, Nyx data uses six 3-D arrays to represent field information in grid structure. 
Both HACC and Nyx data can be mutually verified by each other under the same simulation.
We will discuss more details about the evaluation datasets in Section~\ref{sec:design}.

\section{Problem Statement and Metrics Description}
\label{sec:problem}

In this paper, we focus on quantitatively evaluating and analyzing GPU-based lossy compressors on cosmological simulation datasets. Our evaluation metrics includes the following: 
\begin{enumerate}
\item compression ratio,
\item distortion between original and reconstructed data,
\item cosmological metrics suggested by domain scientists, and
\item compression and decompression throughput.
\end{enumerate}

We now discuss the above metrics in detail.

\textit{Metric 1:} One of the most commonly used metrics in compression research, compression ratio is a ratio of the original data size and the reconstructed data size. Higher compression ratios mean denser information aggregation against the original data, as well as faster data transfer through network or CPU-GPU interconnect. 

\textit{Metric 2}: Distortion is another important metric used to evaluate lossy compression quality in general. In this paper, we use the peak signal-to-noise ratio (PSNR) to measure the distortion quality. Similar to prior work~\cite{tao2017significantly,liangerror,liang2018efficient}, we plot the rate-distortion curve, which is used to compare the distortion quality with the same bitrate (i.e., average number of bits per value), for a fair comparison between different compressors, taking into account diverse compression modes.

\textit{Metric 3}: Besides the above general evaluation metrics, we also take into account cosmology-specific evaluation metrics to better understand the feasibility of using GPU-based lossy compressors in extreme-scale cosmological simulations.
For example, Figs.~\ref{fig:subfigure_orig_crop}, ~\ref{fig:subfigure_sz_rel_error_0.1}, and ~\ref{fig:subfigure_sz_rel_error_0.25} show the visualization of the original and reconstructed Nyx data with GPU-SZ using two different pointwise relative error bounds of 0.1 and 0.25.
Although the two reconstructed data are nearly identical from a visual perspective, the quality of the second one is not acceptable based on a power spectrum analysis as shown in Figure~\ref{fig:psd} (will be discussed in the following section).
In particular, we will analyze the below two cosmology-specific evaluation metrics.

\paragraph{Dark Matter Halos}

Dark matter halos play an important role in the formation and evolution of galaxies and consequently cosmological simulations.
Halos are over-densities in the dark matter distribution and can be identified using different algorithms; in this instance, we use the Friends-of-Friends algorithm~\cite{Davis1985}.
That is, we connect each particle to all ``friends'' within a distance, with a group of particles in one chain considered as one halo. Another concept of halo, such as Most Connected Particle, is defined as the particle within a halo with the most friends. Then there is the Most Bound Particle, which is defined as the particle within a halo with the lowest potential.
A parallel halo-finding function is applied the dataset to generate the results. For decompressed data, some of the information can be distorted from the original.
Information such as the position of one particle can affect the halo number detected, particularly for smaller halos.

\paragraph{Power Spectrum}

Matter distribution in the Universe has evolved to form astrophysical structures on different physical scales, from planets to larger structures such as superclusters and galaxy filaments.
The two-point correlation function $\xi(r)$, which gives the excess probability of finding a galaxy at a certain distance $r$ from another galaxy, statistically describes the amount of the Universe at each physical scale.
The Fourier transform of $\xi(r)$ is called the matter power spectrum $P(k)$, where $k$ is the comoving wavenumber.
Therefore, the matter power spectrum describes how much structure exists at the different physical scales.
Observational data from ongoing sky surveys have measured the power spectrum of matter density fluctuations across several scales.
These sky surveys, along with large-scale simulations, are used to investigate problems such as determining cosmological parameters~\cite{eke2001power}.

\textit{Metric 4:} Compression and decompression throughputs are key advantages of using a GPU-based lossy compressor instead of a CPU-based compressor. 
To make use of GPU's high parallelism---and thus achieve high overall throughput---an efficient GPU kernel is vitally important and must be carefully designed. 
In this paper, we assume that data are generated by cosmological simulations on GPUs, and then the compression would be directly performed on the data from the GPU memory; 
finally, the compressed data would be saved from GPUs to disks via CPUs.
Similarly, we also assume that after the data are reconstructed by GPU decompression, such data would be used by the following simulation or postanalysis tasks on the GPU without transferring it back to the CPU. 
Overall, our goal is to make the compression and decompression throughput as high as possible while still satisfying the postanalysis requirements set by cosmology scientists on the reconstructed data.
\section{Evaluation Methodologies}\label{sec:design}

In this section, we present our evaluation methodologies. We first discuss our proposed Foresight framework. And second, we describe in detail the configurations for the lossy compressors and datasets used in the evaluation. 

\subsection{Design Overview of Foresight}

\begin{figure}[h]
\vspace{-4mm}
\includegraphics[width=8.4cm]{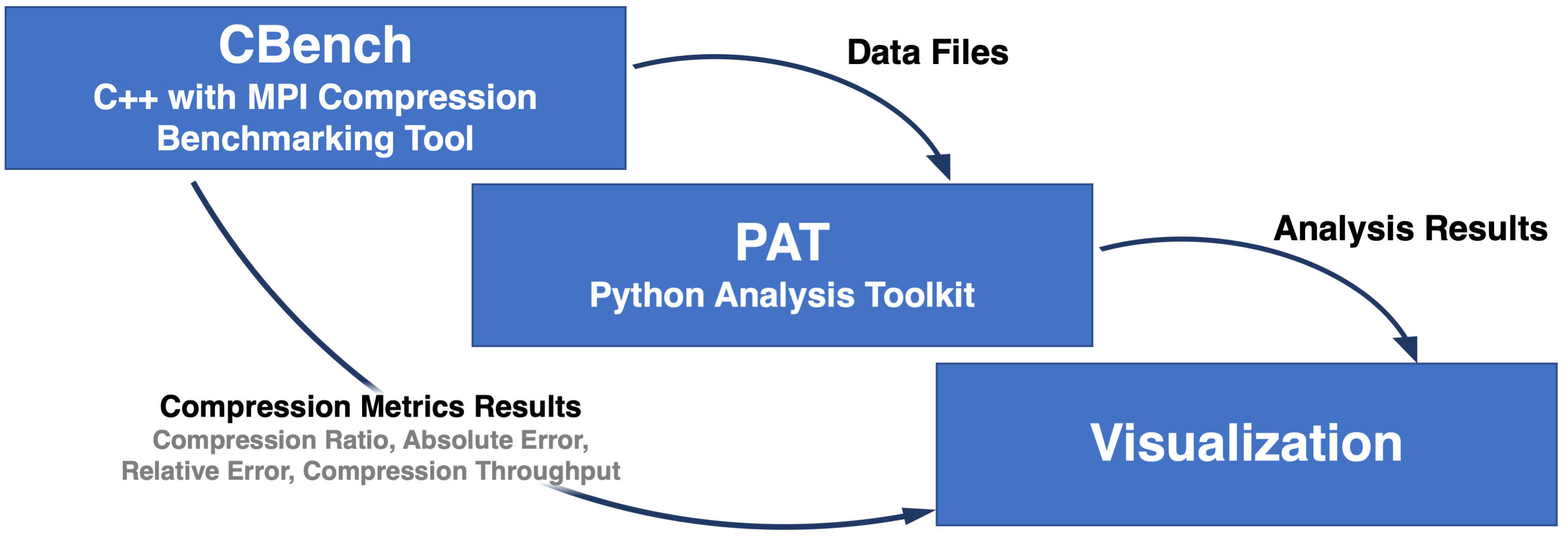}
\centering
\caption{Three components of Foresight framework: CBench executes the compression algorithms, PAT allows CBench to be used in a distributed-computing and \textit{post hoc} analyses, and web-based Cinema viewers are used to visualize the results.}
\label{fig:Components}
\vspace{-3mm}
\end{figure}

We use Foresight\footnote{Available at \href{https://github.com/lanl/VizAly-Foresight}{\color{blue}\underline{https://github.com/lanl/VizAly-Foresight}}.}, an open-source toolkit, as our framework to evaluate, analyze, and visualize lossy compressors for extreme-scale cosmological simulations \cite{foresight-git}, which we enhanced with GPU capabilities.
Foresight is the first toolkit that enables scientists to easily run their own analytics on the compressed data and understand the impacts of lossy compressors on their applications. By only configuring a simple \textit{JSON} file, Foresight can automatically evaluate diverse compression configurations and provide user-desired analysis and visualization on the lossy compressed data. Without Foresight, this process can be achieved only by manually configuring numerous parameters, and it is difficult to identify the bestfit compression solution for different scientific applications and domain-specific metrics.
In addition, we carefully integrate a series of scientific lossy compressors especially GPU-based lossy compressors (including GPU-SZ and cuZFP) to our Foresight framework in this work.
Foresight consists of the following three main components:

\subsubsection{CBench}

CBench is a compression benchmark tool designed for scientific simulations. During recent years, researchers working on extreme-scale scientific simulations have struggled with large data size and have searched for suitable compression methods to shrink the data size for better storage and I/O. 
However, the error generated by error-bounded lossy compression may not always be evenly distributed in the error-bound range.
For instance, lossy compression---such as ZFP---provides a Gaussian-like error distribution.
Moreover, distortion metrics such as MRE, MSE, and PSNR may not have a bijective-function relationship with user-set error bound on the real-world datasets. 
Because of these problems, it is difficult to determine a best-fit configuration (e.g., error bound) for their datasets without any test beforehand. CBench provides researchers with an interface to test different lossy compressors and determine the best-fit compression configuration based on their demands.
The benchmarking results include compression ratio, data distortion (e.g., MRE, MSE, PSNR), compression and decompression throughput, and the reconstructed dataset for the following analysis and visualization in Foresight.

\subsubsection{PAT}

\begin{figure}[]
\vspace{-4mm}
\includegraphics[width=7.8cm]{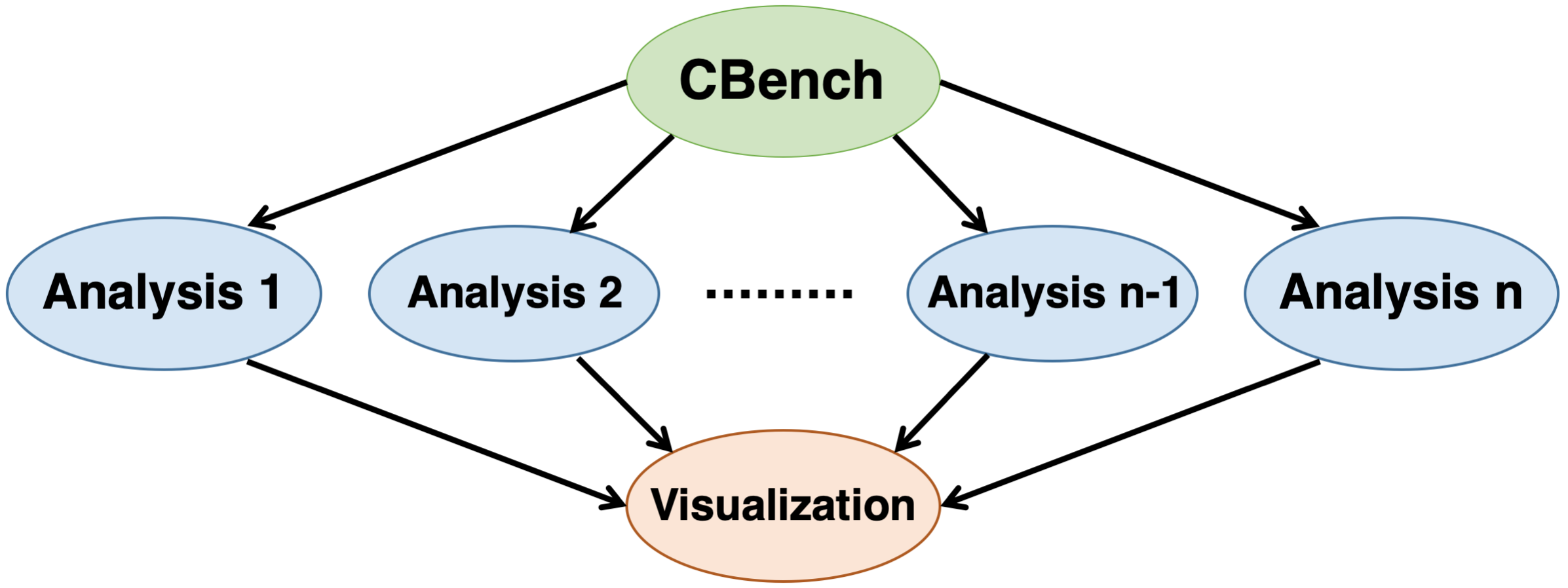}
\centering
\caption{Dependency graph for Foresight.}
\label{fig:Dependency}
\vspace{-8mm}
\end{figure}

Because CBench runs as an executable on the command-line, it does provide one distinguishing feature from other compression evaluation tools: it can perform parameter sweep studies on distributed-computing networks. There are many workflow management systems~\cite{da2017characterization}; however, for the purposes of our study, we require a simple system compatible with the SLURM~\cite{yoo2003slurm} cluster job scheduling system.
Thus, we integrate PAT into our Foresight framework. PAT is a lightweight workflow submission Python package that contains a number of utilities for scheduling SLURM jobs, with dependencies, that execute CBench, provide analysis scripts (\textit{e.g.,} computing power spectra), and plot the results. As shown in Figure~\ref{fig:Components}, PAT takes as input the output files generated by CBench and compression metrics statistics. The two main components of PAT are a \textit{Job} class and a \textit{Workflow} class. The \textit{Job} class enables a user to specify the requirements for a SLURM batch script and the dependencies for that job. The \textit{Workflow} class tracks the dependencies between jobs and writes the submission script for the workflow. The dependency graph for the different components is shown in Figure~\ref{fig:Dependency}.
Note that in this study we adopt cosmology-specific analysis scripts for dark matter halos and power spectrum (discussed in Section \ref{sec:problem}), whereas other analysis code can be added into our framework for different scientific simulations.

\subsubsection{Visualization}
This is the last stage of the pipeline. The \textit{visualization} class will take in metrics from CBench and the files generated from the different analyses by PAT to create plots. The plots are grouped in a Cinema Explorer~\cite{Woodring:2017} database to provide an easily downloadable package for framework users. Examples of Cinema databases can be seen at: \href{https://lanl.github.io/VizAly-Foresight/}{\color{blue}\underline{https://lanl.github.io/VizAly-Foresight/}}. However, in this paper, we only show the plots generated.

\begin{table*}[h]
\centering
\footnotesize
\caption{Specifications of Different GPUs Used in Our Experiments}
\label{tab:gpus}
\renewcommand*{\arraystretch}{1.2}
\footnotesize
\begin{tabular}
{l|ccccccc}
\hline
\multicolumn{1}{c|}{\bfseries GPUs} 
    & \bfseries Release Date 
    & \bfseries Architecture 
    & \bfseries \begin{tabular}{@{}c@{}} \bfseries Compute \\ \bfseries Capability \end{tabular}
    & \bfseries Memory 
    & \bfseries Shaders 
    & \bfseries Peak Perf. (FP32) 
    & \bfseries Memory B/W \\
\hline
    Nvidia RTX 2080Ti & c. 2018 & Turing & \multicolumn{1}{c}{7.5} & \makebox[3.1em][r]{11GB} \makebox[4.3em][l]{GDDR6}& \multicolumn{1}{c}{4352} & \makebox[2.7em][r]{13} \makebox[4em][l]{TFLOPS} & \makebox[3.5em][r]{448} \makebox[3.2em][l]{GB/s} \\
    Nvidia Tesla V100 & c. 2017 & Volta & 7.0-7.2 & \makebox[3.1em][r]{16GB} \makebox[4.3em][l]{HBM2}& \multicolumn{1}{c}{5120} & \makebox[2.7em][r]{14} \makebox[4em][l]{TFLOPS} & \makebox[3.5em][r]{900} \makebox[3.2em][l]{GB/s} \\
    Nvidia Titan V & c. 2017 & Volta & 7.0-7.2 & \makebox[3.1em][r]{12GB} \makebox[4.3em][l]{HBM2}& \multicolumn{1}{c}{5120} & \makebox[2.7em][r]{15} \makebox[4em][l]{TFLOPS} & \makebox[3.5em][r]{650} \makebox[3.2em][l]{GB/s} \\
    Nvidia GTX 1080Ti & c. 2017 & Pascal & 6.0-6.2 & \makebox[3.1em][r]{11GB} \makebox[4.3em][l]{GDDR5X}& \multicolumn{1}{c}{3584} & \makebox[2.7em][r]{11} \makebox[4em][l]{TFLOPS} & \makebox[3.5em][r]{485} \makebox[3.2em][l]{GB/s} \\
    Nvidia P6000 & c. 2016 & Pascal & 6.0-6.2 & \makebox[3.1em][r]{24GB} \makebox[4.3em][l]{GDDR5X}& \multicolumn{1}{c}{3840} & \makebox[2.7em][r]{13} \makebox[4em][l]{TFLOPS} & \makebox[3.5em][r]{433} \makebox[3.2em][l]{GB/s} \\
    Nvidia Tesla P100 & c. 2016 & Pascal & 6.0-6.2 & \makebox[3.1em][r]{16GB} \makebox[4.3em][l]{HBM2}& \multicolumn{1}{c}{3584} & \makebox[2.7em][r]{9.5} \makebox[4em][l]{TFLOPS} & \makebox[3.5em][r]{732} \makebox[3.2em][l]{GB/s} \\
    Nvidia Tesla K80 & c. 2014 & Kepler 2.0 & 3.0-3.7 & \makebox[3.1em][r]{12$\times$2GB} \makebox[4.3em][l]{GDDR5}& 2496$\times$2 & \makebox[2.7em][r]{4$\times$2} \makebox[4em][l]{TFLOPS} & \makebox[3.5em][r]{240$\times$2} \makebox[3.2em][l]{GB/s} \\
\hline
\end{tabular}%

\vspace{-5mm}
\end{table*}

\subsection{Evaluation Configuration}

\subsubsection{Lossy Compressors}

Many recent studies~\cite{tao2018optimizing,lu2018understanding,luo2019identifying} have showed that SZ and ZFP are two leading lossy compressors for scientific simulation data and can well represent prediction- and transformation-based lossy compressors, respectively.
In this paper, we mainly focus on these two advanced lossy compressors without loss of generality.

SZ has implemented a GPU version prototype using OpenMP (denoted by GPU-SZ~\cite{sz-openmp}). However, the memory layout implemented in the current version is not optimized for overall throughput. Based on our discussion with the SZ development team, the compression and decompression throughputs are expected to be significantly improved after the memory-layout optimization, which can also maintain the same compression quality. Therefore, in this paper we evaluate all metrics except the compression and decompression throughput for GPU-SZ.  
We also note that the current GPU-SZ only supports 3-D dataset and absolute error-bound mode (\texttt{ABS}). Accordingly, for the 1-D HACC dataset, we will first convert it into 3-D format and second compress the converted 3-D data. 
It is worth noting that the point-wise relative error-bound mode (\texttt{PW\_REL}) can result in better compression quality for some data fields, such as HACC velocity data. 
In this case, we will convert the original data by a logarithmic transformation and then compress the converted data using the ABS mode, which can achieve the same effect as \texttt{PW\_REL} mode, according to an existing work~\cite{liang2018efficient}. 

ZFP has released its CUDA version named cuZFP~\cite{cuZFP}. We note that the current cuZFP only supports compression and decompression with fixed-rate mode. Thus, we will use cuZFP with different bitrates in our evaluation.

\begin{table}[h]
\vspace{1.5mm}
\renewcommand*{\arraystretch}{1.3}
\centering
\footnotesize
\caption{Details of HACC and Nyx Dataset Used in Experiments}
\vspace{-1mm}
\begin{tabular}{r|c|c|c|c}
\hline
\textbf{Dataset}      
	& \textbf{Dimension}                         
	& \textbf{Size}          
	& \textbf{Field}
	& \textbf{Value Range} 
	\\ \hline
\multicolumn{1}{c|}{\multirow{2}{*}{HACC}}
	& \multirow{2}{*}{1,073,726,359}             
	& \multirow{2}{*}{38 GB}  
	& Position 
	& $(0, 256)$         
	\\ \cline{4-5} 
	&                                            
	&                        
	& Velocity 
	& $\left(-10^4, 10^4\right)$    
	\\ \hline
\multicolumn{1}{c|}{\multirow{4}{*}{Nyx}}  
	& \multirow{4}{*}{512$\times$512$\times$512} 
	& \multirow{4}{*}{6.6 GB} 
	& Baryon Density
	& $\left(0, 10^5\right)$        
	\\ \cline{4-5} 
	&                                            
	&                        
	& {\renewcommand*{\arraystretch}{1.05}
	\begin{tabular}[c]{@{}c@{}}Dark Matter \\ Density\end{tabular} }
	& $\left(0, 10^4\right)$        
	\\ \cline{4-5} 
	&                                            
	&                        
	& Temparature                                                    
& $\left(10^2, 10^7\right)$     
	\\ \cline{4-5} 
	&                                            
	&                        
	& Velocity
	& $\left(-10^8, 10^8\right)$    
	\\ \hline
\end{tabular}
\vspace{-4.5mm}
\label{tab:DataDetail}
\end{table}

\subsubsection{Evaluation Datasets}

The cosmological simulation data used in our experimental evaluation includes both HACC and Nyx datasets. The details of the datasets are shown in Table~\ref{tab:DataDetail}.

The HACC dataset\footnote{Available at \href{http://dx.doi.org/10.21227/zg3m-8j73}{\color{blue}\underline{http://dx.doi.org/10.21227/zg3m-8j73}}.} is provided by the HACC development team at Argonne National Laboratory \cite{zg3m-8j73-19}. It contains six 1-D arrays, with each field having 1,073,726,359 single-precision floating-point numbers. The HACC simulation used to generate this dataset uses module M001 to cover a $(0.36~\text{Gpc})^3$ volume and sets the redshift value to be 0. The data are in GenericIO file format~\cite{genericio}. 
The Nyx dataset\footnote{Available at \href{http://dx.doi.org/10.21227/k8gb-vq78}{\color{blue}\underline{http://dx.doi.org/10.21227/k8gb-vq78}}.} is provided by the Nyx development team at Lawrence Berkeley National Laboratory \cite{k8gb-vq78-19}. It is a single-level grid structure without adaptive mesh refinement (AMR) and does not include particle data. It contains six 3-D arrays, with each field having 512$\times$512$\times$512 single-precision floating-point numbers. The six fields are baryon density $(\rho_b)$, dark matter density $(\rho_d{}_m)$, temperature
$(T)$, and velocity in three directions $(v_x, v_y, v_z)$. The dataset is in HDF5 file format~\cite{folk1999hdf5}.

\subsubsection{Experimental Platform}

We conduct our evaluation on two HPC clusters, which are the Darwin~\cite{darwin} cluster at Los Alamos National Laboratory and PantaRhei~\cite{pantarhei} cluster at the University of Alabama. We perform our experiments on different GPUs as shown in Table~\ref{tab:gpus} from both clusters and compare with lossy compressors on 20-core Intel Xeon Gold 6148 CPU from PantaRhei. All the GPUs are connected to the host via 16-lane PCIe 3.0 interconnect. 

\subsubsection{Implementation Details}
\hfill\\
\indent\textbf{Dimension conversion:}
As we mentioned above, the current GPU-SZ does not support 1-D dataset. 
To compress HACC 1-D dataset properly, we need to convert the six 1-D arrays into 3-D arrays. 
In fact, we must convert the 1-D HACC dataset to 3-D because the HACC simulation used to generate this dataset runs with 8$\times$8$\times$4 MPI processes, and each MPI process saves its own portion of the dataset, leading to 8$\times$8$\times$4 data partitions.
We first divide the 1,073,726,359 data points of each 1-D array into eight 134,217,728 $\left(2^{27}\right)$ partitions, padding with zeros. Then, we convert each 1-D partition into 512$\times$512$\times$512 or 2,097,152$\times$8$\times$8 size of 3-D arrays. Consequently, we can use GPU-SZ to compress the converted 3-D data. 
We note that the 512$\times$512$\times$512 conversion results in best compression quality in our experiments. Thus, we will use 512$\times$512$\times$512 for GPU-SZ in the following evaluation.
Although cuZFP can support 1-D data, the compression quality on the 1-D data is not as good as that on the converted 3-D data. 
Accordingly, we use the 3-D size of 2,097,152$\times$8$\times$8 for cuZFP on the HACC dataset. 
This process will be reversed for reconstruction data.
We decompress the compressed 3-D dataset and then convert the decompressed dataset back to 1-D arrays.
The time overhead of this conversion is negligible because we only pass the pointer and specify the data dimension to the compression and decompression function. 

\textbf{Logarithmic transformation:}
Based on a previous study~\cite{liang2018efficient} and a suggestion from the HACC developers, \texttt{PW\_REL} is better than ABS for the velocity fields in the HACC dataset. However, GPU-SZ only supports ABS so far.
Inspired by a previous work~\cite{liang2018efficient}, we adopt a logarithmic transformation with the base $e$ to convert each velocity field point by point and the user-set \texttt{PW\_REL} error bound to ABS error bound.
We then perform the GPU-SZ compression on the converted data and ABS error bound. 
During the reconstruction, we decompress the data using the converted ABS error bound and then reconstruct the data with the exponential transformation using the same base $e$.

\section{Experimental Results and Analysis}
\label{sec:evaluation}

In this section, we first present the evaluation results\footnote{We provide the detailed instructions for readers to reproduce our experimental results at \href{https://github.com/jinsian/VizAly-Foresight}{\color{blue}\underline{https://github.com/jinsian/VizAly-Foresight}}.} of different GPU-based lossy compressors on the HACC and Nyx datasets and second provide a general optimization guideline for cosmology researchers and scientists.  

\subsection{Rate-Distortion Evaluation}
\label{sec:rate-distortion-evaluation}

\begin{figure}[]
\begin{subfigure}{0.92\linewidth}\centering
    \includegraphics[width=\linewidth]{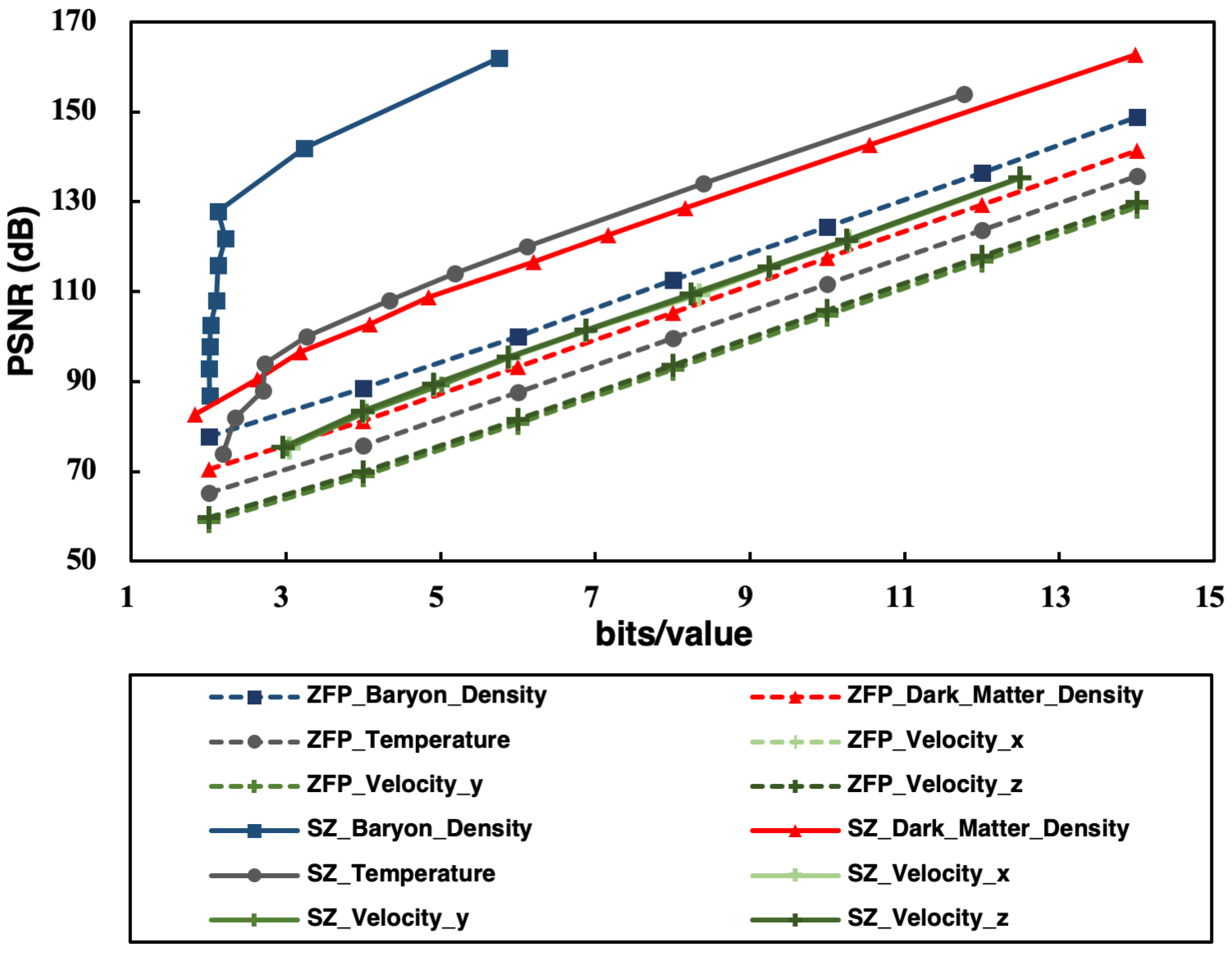}
	\caption{\footnotesize Nyx}\label{subfig:rate-distortion_a}
\end{subfigure}
\begin{subfigure}{0.92\linewidth}\centering
    \includegraphics[width=\linewidth]{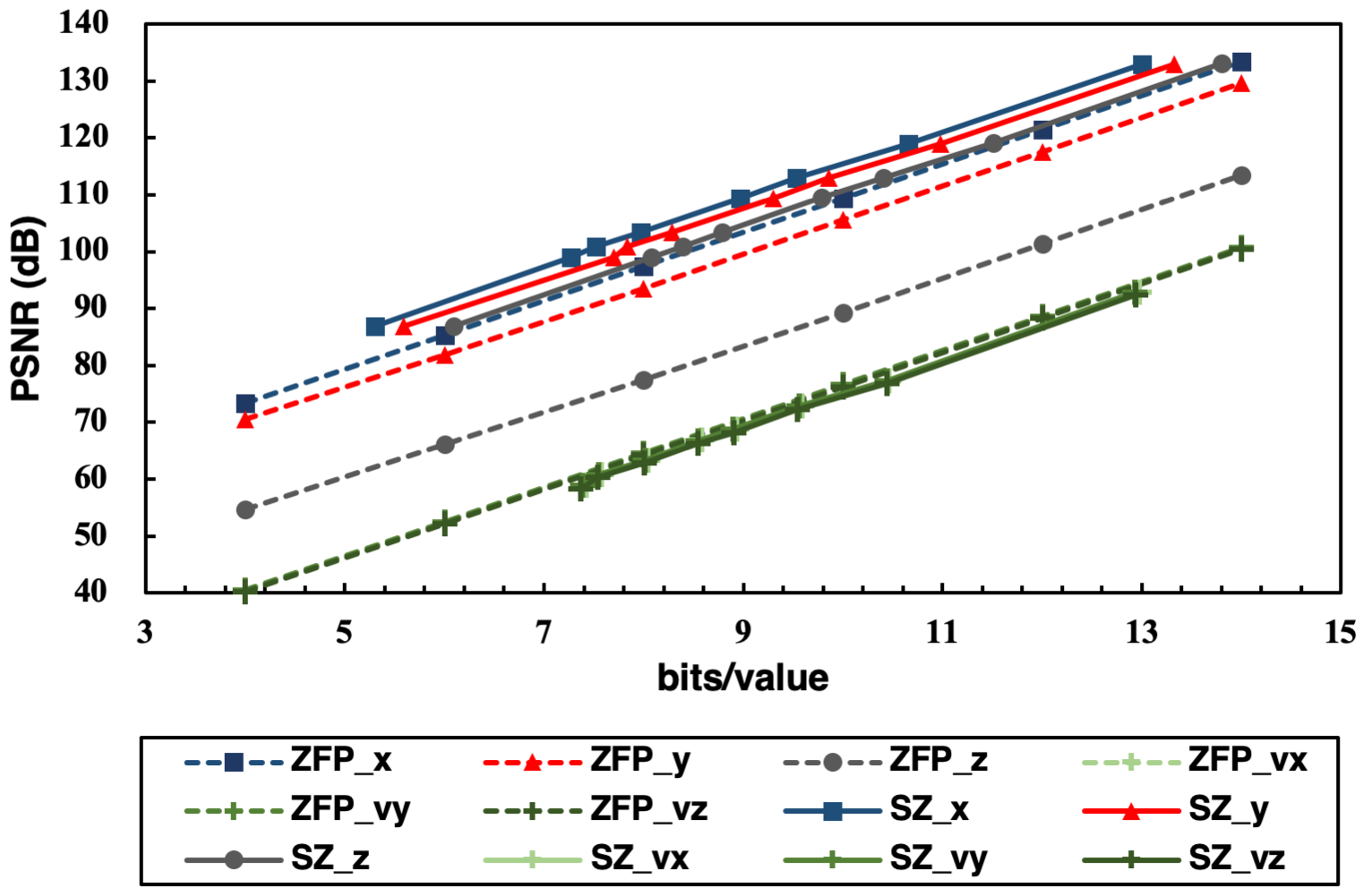}
	\caption{\footnotesize HACC}\label{subfig:rate-distortion_b}
\vspace{-1mm}
\end{subfigure}
\caption{Rate-distortion of GPU-SZ and cuZFP on HACC and Nyx datasets. \small{Solid lines represent compression by GPU-SZ. Dashed lines represent compression by cuZFP.}}
\label{fig:rate-distortion}
\vspace{-8mm}
\end{figure}

\begin{figure*}[]
\begin{subfigure}{\linewidth}\centering
    \includegraphics[width=17.2cm]{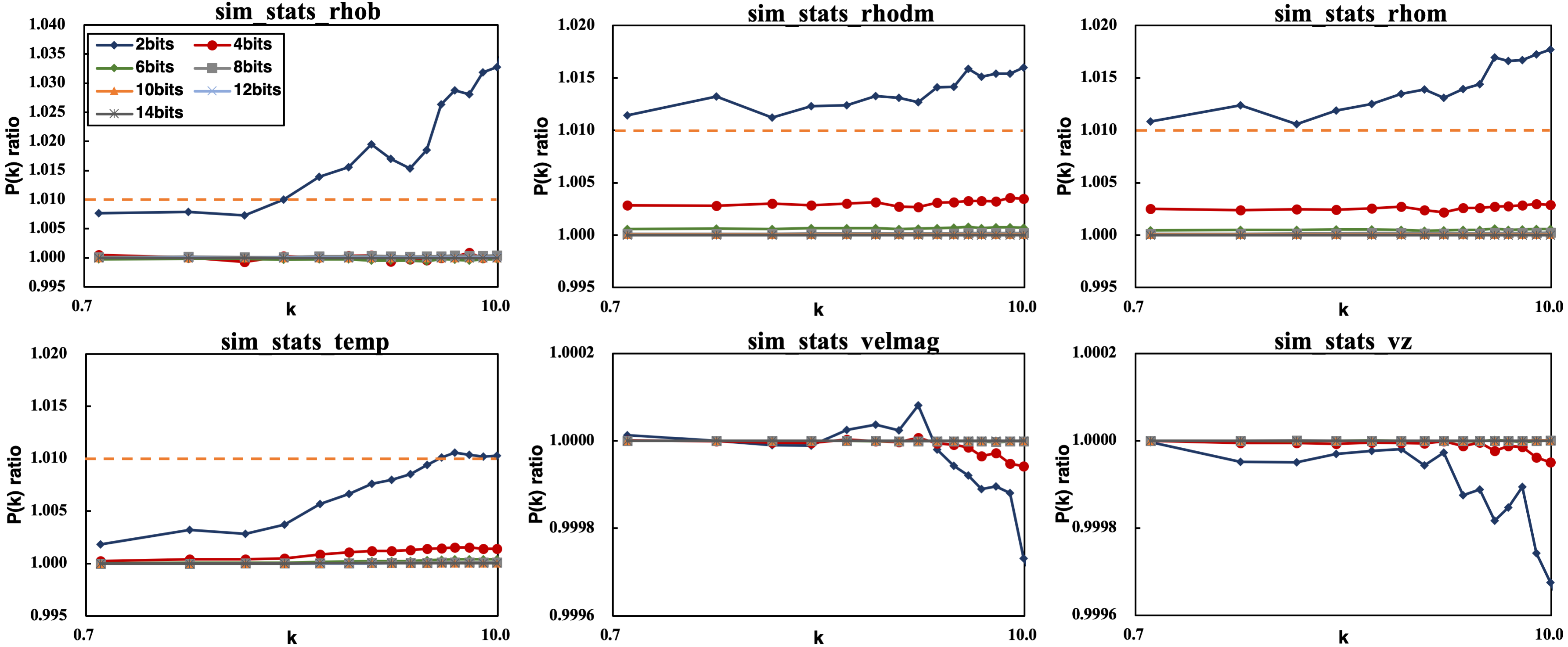}
    \vspace{-2mm}
	\caption{\footnotesize cuZFP}\label{subfig:nyx-zfp-pwer_a}
\end{subfigure}
\begin{subfigure}{\linewidth}\centering
    \includegraphics[width=17.2cm]{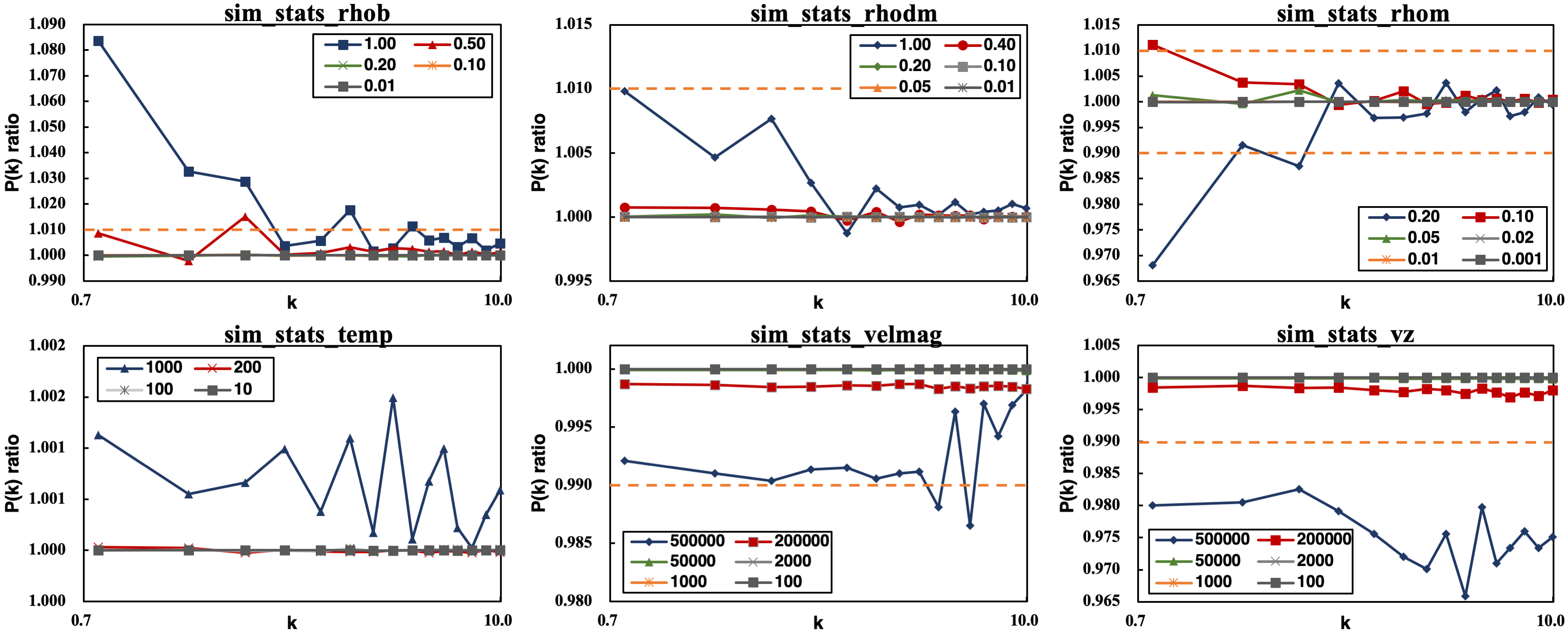}
    \vspace{-2mm}
	\caption{\footnotesize GPU-SZ}\label{subfig:nyx-zfp-pwer_b}
\end{subfigure}
\caption{Power spectrum analysis on different reconstructed fields in Nyx dataset using cuZFP and GPU-SZ with different bitrates and ratios, respectively. The orange dashed line stands for a pk ratio constraint of 1$\pm$1\%.}
\label{fig:nyx-zfp-pwer}
\vspace{-6mm}
\end{figure*}

We first present the results of compression ratio and data distortion as a result of using GPU-SZ and cuZFP on our tested cosmological simulation datasets. We use PSNR to evaluate the quality of the reconstructed data. Note that GPU-SZ is designed for fixed maximum compression error using its \texttt{ABS} mode, whereas cuZFP is designed for fixed compression ratio using its fixed-bit mode. For a fair comparison, we plot the rate-distortion curves for both GPU-SZ and cuZFP on different fields and compare their distortion values in PNSR at the same bitrate, as shown in Figure~\ref{fig:rate-distortion}. The $x$-axis stands for the bitrate in bits/value. The lower a bitrate is, the higher its compression ratio will be. 
Note that the original HACC and Nyx data is in 32-bit single precision, meaning that a bitrate of 4.0 is equivalent to the compression ratio of 8$\times$.

Figure~\ref{subfig:rate-distortion_a} shows the rate-distortion curve of GPU-SZ and cuZFP on the Nyx dataset. 
We note that GPU-SZ generally has higher PSNR than cuZFP with the same bitrate on the Nyx dataset. In other words, GPU-SZ can provide a higher compression ratio compared to cuZFP, given the same reconstructed data quality. 
This can be attributed to (1) the \texttt{ABS} mode having better compression performance than the fixed-rate mode on the Nyx dataset (cuZFP has not supported the \texttt{ABS} mode yet), and (2) the adaptive predictor \cite{liangerror} (Lorenzo or regression-based predictor) of GPU-SZ has higher decorrelation efficiency than the block transform of cuZFP on the field with extremely large value range but concentrated distribution, such as the fields in the Nyx dataset.
Furthermore, we observe that most of the rate-distortion curves linearly increase with the bitrate and have similar slopes. However, the rate-distortion curves of GPU-SZ on baryon density and temperature field drop dramatically when the bitrate is lower than about 2. 
Compared with the results shown in the previous work~\cite{liangerror}, we infer that this drop could be caused by the GPU-SZ dataset blocking, which divides the data into multiple independent blocks and decorrelates at the block borders, leading to more unpredictable data points and a lower compression ratio, given the same level of distortion.
Note that the compression quality of cuZFP with block transform degrades noticeably on baryon density because it has relatively more extreme values compared to the other fields in the Nyx dataset.
Lastly, we observe that for both GPU-SZ and cuZFP, their rate-distortion curves for velocity fields are almost identical. This is because velocity fields have similar data characteristics, which is more randomly compared with the other fields, according to previous studies~\cite{tao2017depth,li2018optimizing}.

Figure~\ref{subfig:rate-distortion_b} shows the rate-distortion curves of GPU-SZ and cuZFP on the HACC dataset. Similar to previous observations, we note that the linearity between PSNR and bitrate for both GPU-SZ and cuZFP. Compared to cuZFP, GPU-SZ has comparable performance on the velocity fields $(v_x, v_y, v_z)$ but better performance on the position fields $(x, y, z)$. This is because we adopt the \texttt{PW\_REL} mode on the velocity data for GPU-SZ.
Unlike \texttt{ABS} mode, which introduces the same level of distortion to each data point, \texttt{PW\_REL} mode introduces the error based on the data point value.
It is worth noting that (1) PSNR is sensitive to the large error given the same bitrate, and (2) higher PSNR does not necessarily indicate better postanalysis quality, especially when taking into consideration the cosmology-specific metrics such as power spectrum and dark matter halos. 
In our experiments, we observe that the \texttt{PW\_REL} mode can result in better compression quality in terms of power spectrum and halo finder (illustrated in the next section). Thus, we use in this paper the \texttt{PW\_REL} mode for the velocity fields in the HACC dataset.

\subsection{Power Spectrum and Halo Finder Analysis}

Next, we evaluate the reconstructed data based on power spectrum and halo finder. The analysis is generated by Foresight's PAT and Cinema.

Figure~\ref{subfig:nyx-zfp-pwer_a} shows six power spectra of the Nyx dataset with different compression ratios. 
Note that the power spectra are not in a one-to-one mapping to each data field; instead, some power spectra are composites of two or more fields.
Specifically, the sub-figures from left to right and top to bottom are the power spectra on baryon density, dark matter density, overall density (baryon density and dark matter density), temperature, velocity magnitude~($\sqrt{vx^2+vy^2+vz^2}$), and velocity $vz$. 
Note that we only show the power spectrum of velocity $vz$ because it is similar to that of velocities $vx$ and $vy$. Here the $x$-axis stands for frequency in log scale, and the $y$-axis is the ratio of the power spectra on the reconstructed data and that on the original data (pk ratio).
Our primary goal is to identify a compression configuration that can maintain the pk ratio within 1$\pm$1\% and has the highest compression ratio. We will further explain the reason for choosing the highest compression ratio in the following discussion.
From the figure, we observe that when the bitrate is 2, except for velocity fields, the pk ratios of the remaining fields are lower than 0.99 or larger than 1.01, which is unacceptable.
Therefore, we choose the fixed bitrate of $(4, 4, 4, 2, 2, 2)$ for the six fields, respectively, with cuZFP, which will result in an overall compression ratio of 10.7$\times$. 
Similarly, based on Figure~\ref{subfig:nyx-zfp-pwer_b}, we can with GPU-SZ determine the error bound of $(0.2, 0.4, 1\text{e+}3, 2\text{e+}5, 2\text{e+}5, 2\text{e+}5)$ for the six fields, respectively. The overall compression ratio under this configuration is 15.4$\times$.

As we mentioned in the previous discussion, higher PSNR does not necessarily indicate better compression quality in terms of power spectrum and halo finder. 
Compared to the result shown in Section~\ref{sec:rate-distortion-evaluation},
cuZFP can maintain the pk ratio within 1$\pm$1\% using the bitrate of 4 for baryon density, and its corresponding PSNR is 88.45 dB. However, GPU-SZ using the absolute error bound of 1.0 has the PSNR of 102.45 dB. However, its pk ratio exceeds 1.01, which is unacceptable.
In fact, when choosing the absolute error bound of 0.2 for baryon density, GPU-SZ can provide acceptable pk ratio and higher compression ratio than cuZFP.
The above observations demonstrate the necessity of domain-specific metrics evaluation rather than only the general rate-distortion evaluation.

\begin{figure}[h]
\begin{subfigure}{\linewidth}\centering
    \includegraphics[width=8.8cm]{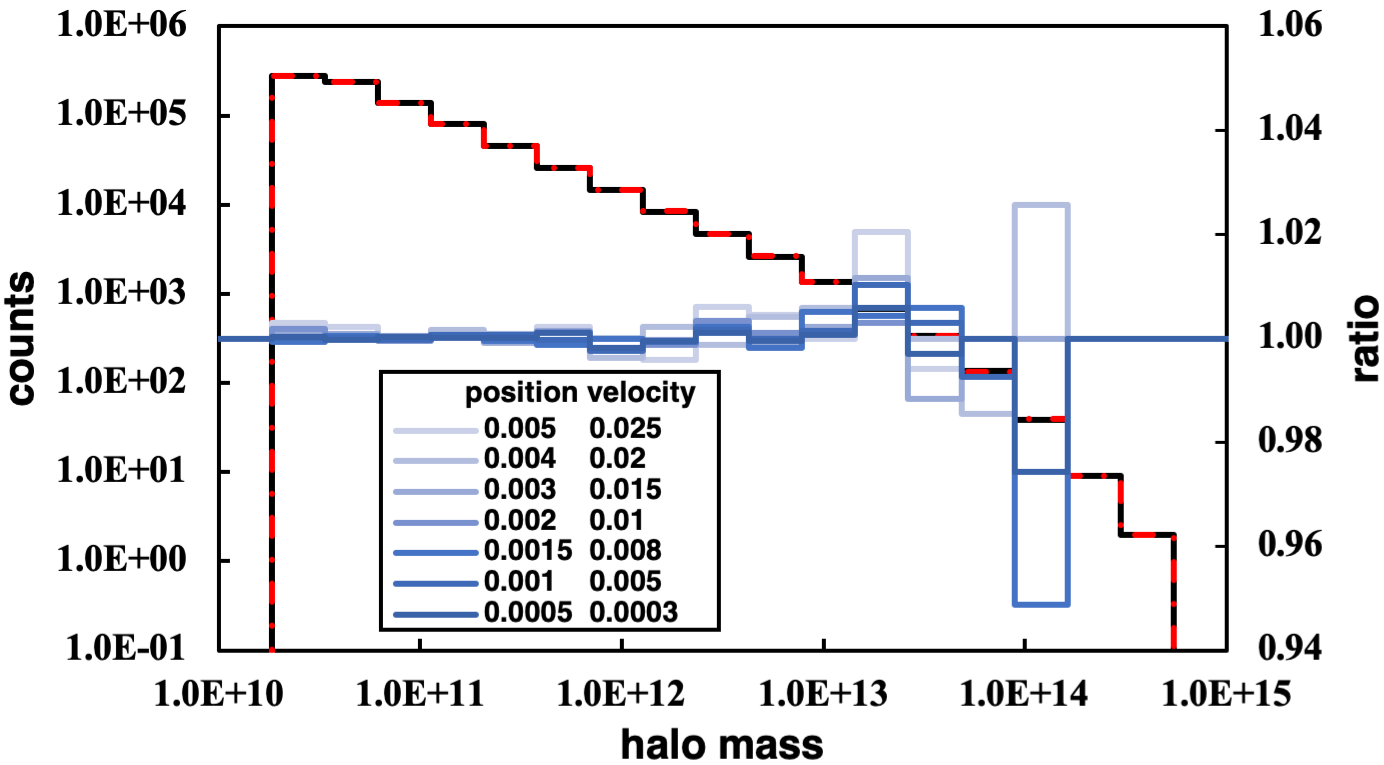}
	\caption{\footnotesize GPU-SZ}\label{subfig:halo-finder_a}
\end{subfigure}
\begin{subfigure}{\linewidth}\centering
    \includegraphics[width=8.8cm]{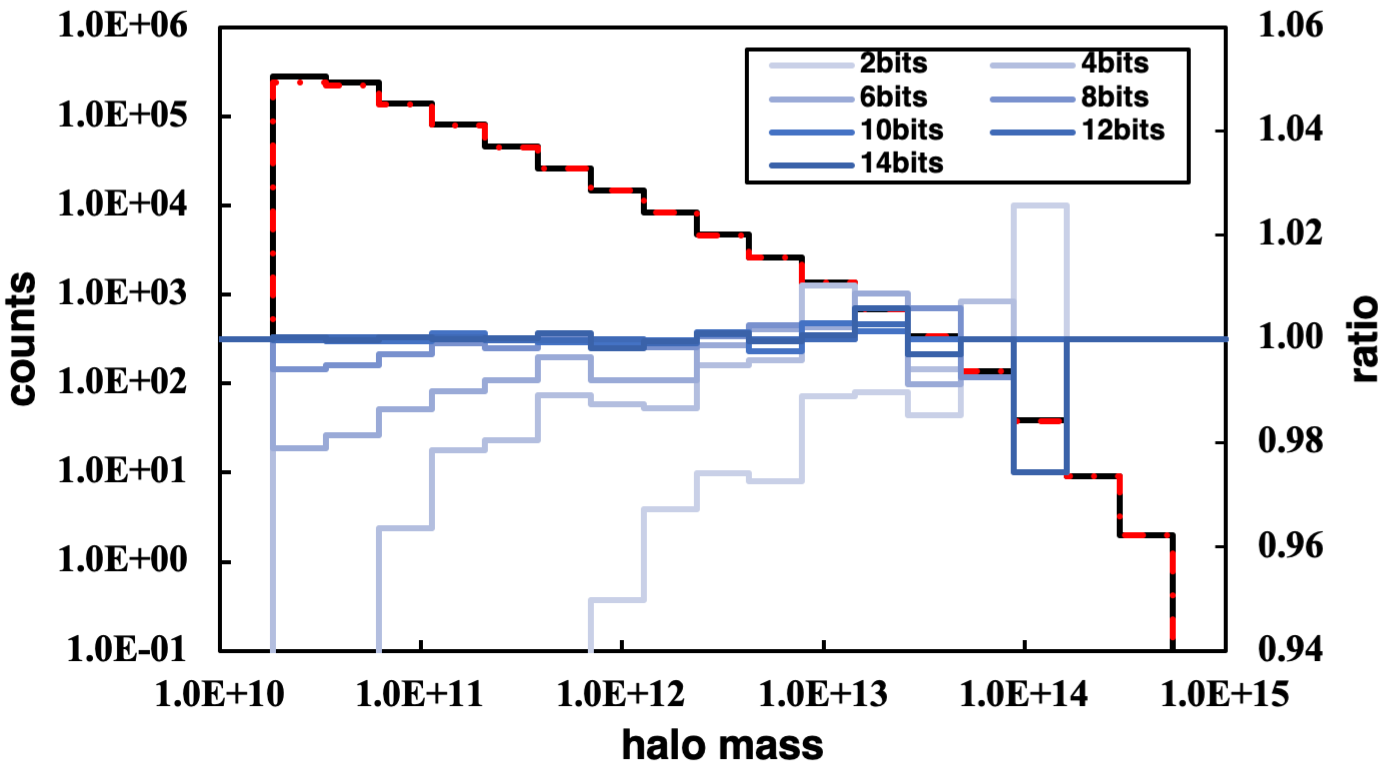}
	\caption{\footnotesize cuZFP}\label{subfig:halo-finder_b}
\end{subfigure}
\caption{Comparison of halo finder analysis on original and reconstructed HACC dataset. \small{X-axis represents the halo mass. Left y-axis represents the counts of halos (black line is for original data, red dashed line is for reconstructed data). Right y-axis represents the ratio of counts generated by reconstructed data and original data.}}
\label{fig:halo-finder}
\vspace{-6mm}
\end{figure}

For the HACC dataset, Figure~\ref{fig:halo-finder} illustrates the comparison of the halo finder analysis on the original data and reconstructed data with GPU-SZ and cuZFP.
We observe that the counts of halos generated by the original data (black line) and reconstructed data (dashed red line) are close to each other. 
We also plot the curves for the ratio of the halo counts on the original data and the halo counts on the reconstructed data with different compression ratios, as shown in Figure~\ref{subfig:halo-finder_a}.
It is worth noting that all the halo count ratios of GPU-SZ are close to 1, which demonstrates that the halo information is well preserved during the lossy compression. Therefore, we choose the absolute error bound of 0.005 and 0.025 for the position and velocity fields, respectively, leading to the highest compression ratio of $4.25\times$.
To maintain a high accuracy of halo finder analysis, we must choose the fixed bitrate higher than 8 for cuZFP, as shown in Figure~\ref{subfig:halo-finder_b}. This configuration results in an overall compression of 4$\times$, which is slightly lower than GPU-SZ's $4.25\times$. Note that we use different compression modes for GPU-SZ and cuZFP. 

\subsection{Throughput Evaluation}

In this section, we discuss our evaluation on the GPU-based lossy compression throughput based on the tested datasets.
For the time measurement of each CUDA (de)compression kernel, we first warm up the GPU device by running the kernel 10 times, and then we measure the times of next 10 runs and calculate the average and standard deviation. 
For measuring the time of single-core or multi-core (de)compression, we perform each experiment 5 times and calculate the average and standard deviation.
We observe that all the standard deviation values are relatively negligible compared to their average values because our experiments are conducted under a very stable environment by dominating the entire GPU nodes.

\begin{figure}[]
\begin{subfigure}{\linewidth}\centering
    \includegraphics[width=8.8cm]{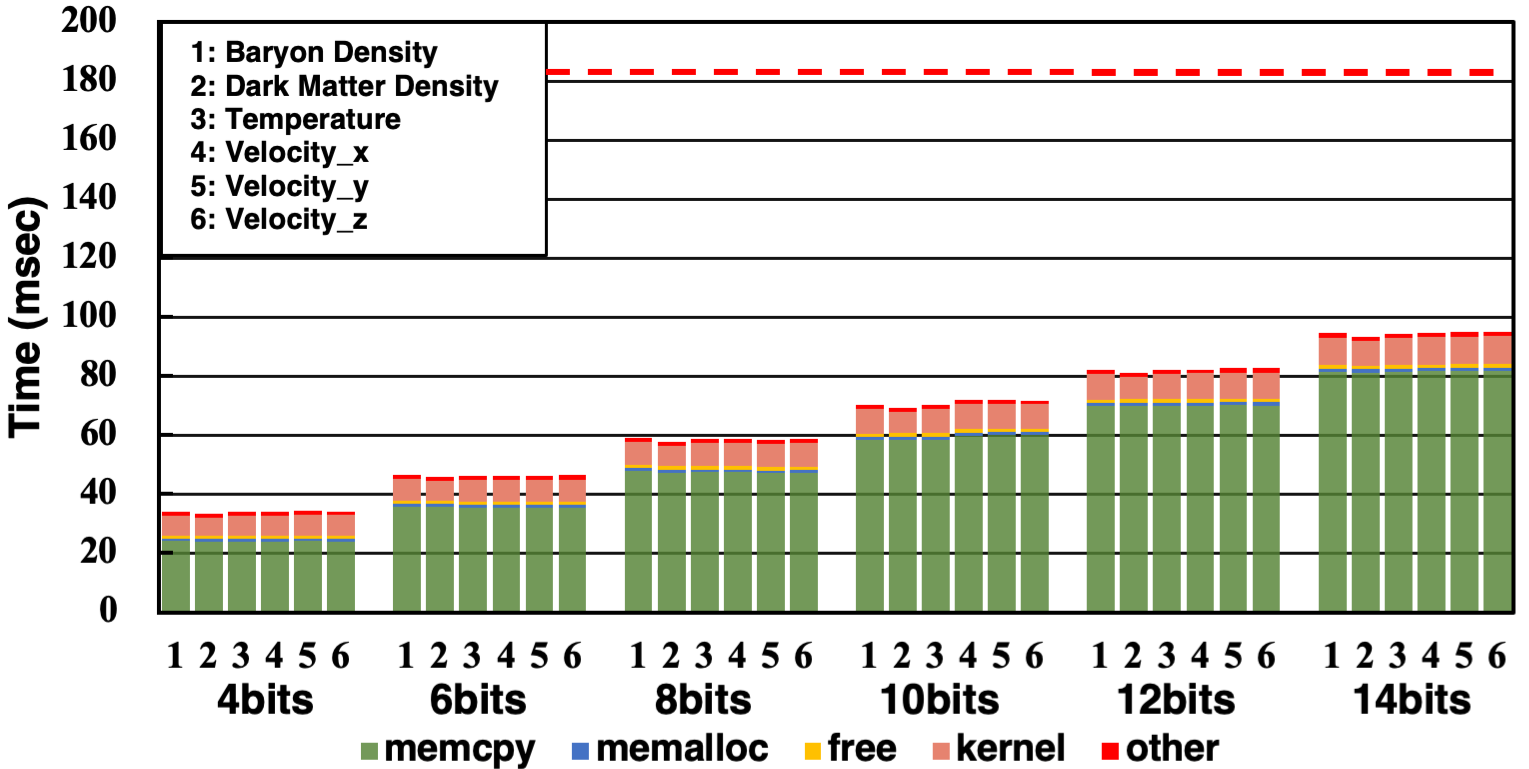}
	\caption{\footnotesize compression}\label{subfig:breakdown_a}
\end{subfigure}
\begin{subfigure}{\linewidth}\centering
    \includegraphics[width=8.8cm]{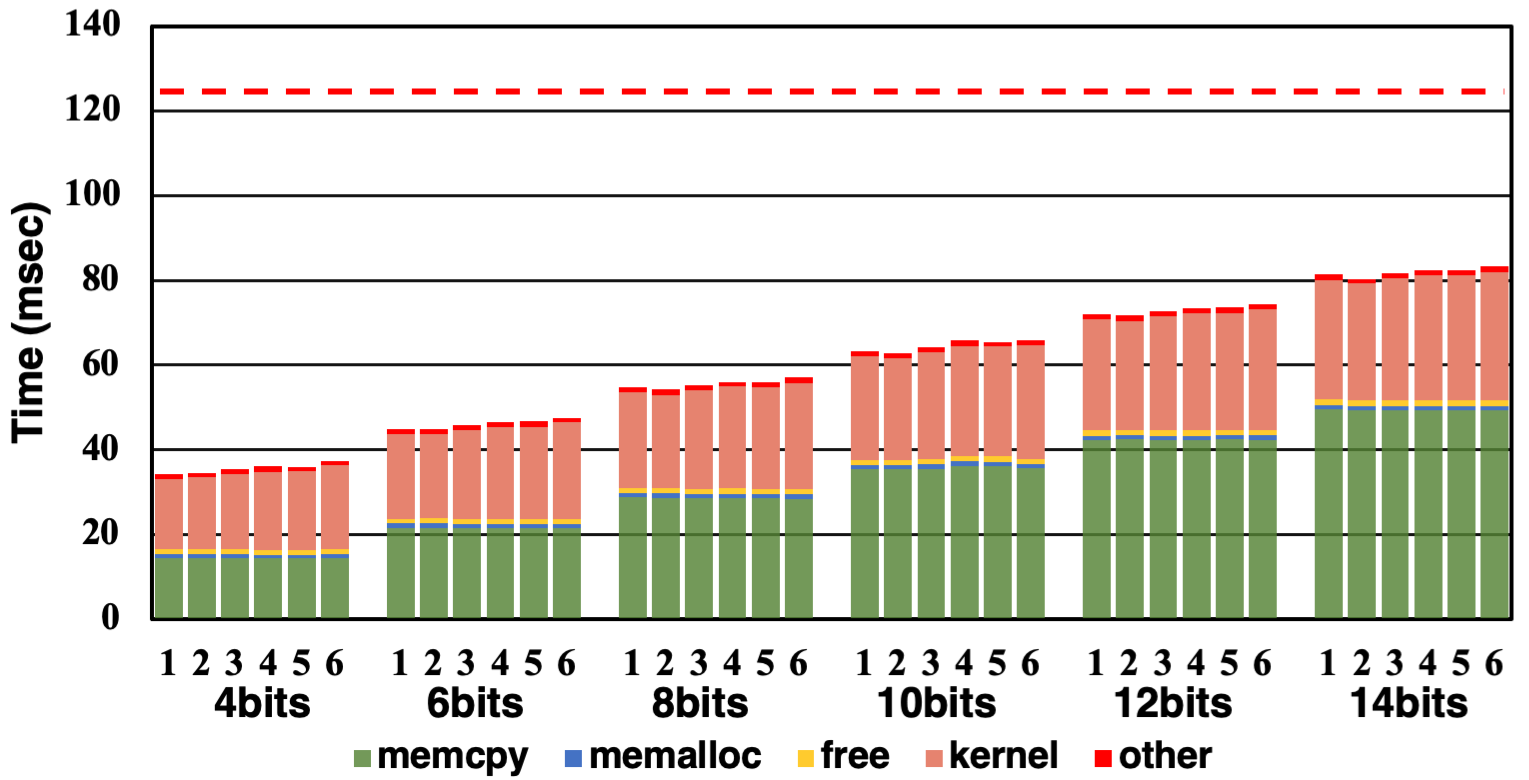}
	\caption{\footnotesize decompression}\label{subfig:breakdown_b}
	\vspace{-2mm}
\end{subfigure}
\caption{Breakdown of compression and decompression time with cuZFP on Nyx dataset.}
\label{fig:breakdown}
\vspace{-8mm}
\end{figure}

Throughput is one of the most important advantages for using GPU-based lossy compression over traditional CPU-based compression. Before comparing the overall GPU and CPU throughput, we must understand how much time each GPU-based lossy compression and decompression component spends. 
As discussed in Section~\ref{sec:problem}, we assume that the original data are stored in the GPU memory after generation, and then the compressed data would be moved from GPU to disk through CPU after compression.
Figure~\ref{subfig:breakdown_a} shows the time breakdown of compression with cuZFP on the Nyx dataset using different bitrates. cuZFP compression and decompression contain the following steps: 
(1) copying the compression parameters from CPU to GPU and preallocating the GPU memory space for the compressed or decompressed data, denoted by \texttt{init} (i.e., blue bar);
(2) launching the compression kernel and performing compression or decompression, denoted by \texttt{kernel} (i.e., pink bar);
(3) copying the compressed data back to CPU, denoted by \texttt{memcpy} (i.e., green bar); and
(4) deallocating the GPU memory space, denoted by \texttt{free} (i.e., yellow bar).
The red dashed line in Figure~\ref{subfig:breakdown_a} is the baseline of copying the original data from GPU to CPU without any compression.

From Figure~\ref{subfig:breakdown_a}, we observe that given the same bitrate, the overall compression and decompression time and their breakdown are relatively consistent across different data fields. The reason is that the most time-consuming step \texttt{memcpy} and \texttt{kernel} are directly related to the data size, whereas the same bitrate will lead to the same data size. 
We also note that the compression and decompression time of higher bitrate is longer than that of lower bitrate. It is because \texttt{memcpy} (i.e., GPU-CPU in compression or CPU-GPU in decompression) takes more time for the compressed data with higher user-set bitrate.
In addition, the kernel throughput is also decreased by increasing the bitrate. This is important for optimizing the compression configuration for cosmological simulations that will be discuss more in the next section. 
It is worth noting that the compression kernel time on GPU is relatively low compared to the data transfer time between GPU and CPU (i.e., \texttt{memcpy}). However, it can be significantly reduced by leveraging new CPU-to-GPU interconnect technology with higher bandwidth such as NVLink or asynchronous GPU-CPU communication.
Therefore, the high-throughput GPU (de)compression introduces only a negligible overhead to the overall performance of the simulation.

Figure~\ref{subfig:breakdown_b} shows the time breakdown of decompression with cuZFP. Similarly, we assume that the compressed data must be copied from CPU to GPU and then decompressed with the GPU kernel; and the decompressed data will be directly passed to the following simulation or analysis tasks. The observations here are similar to compression, and the main performance bottleneck is the data transfer time from CPU to GPU, which demonstrates that higher compression ratio is important to the overall (de)compression throughput. 

\begin{figure}[]
\includegraphics[width=8.8cm]{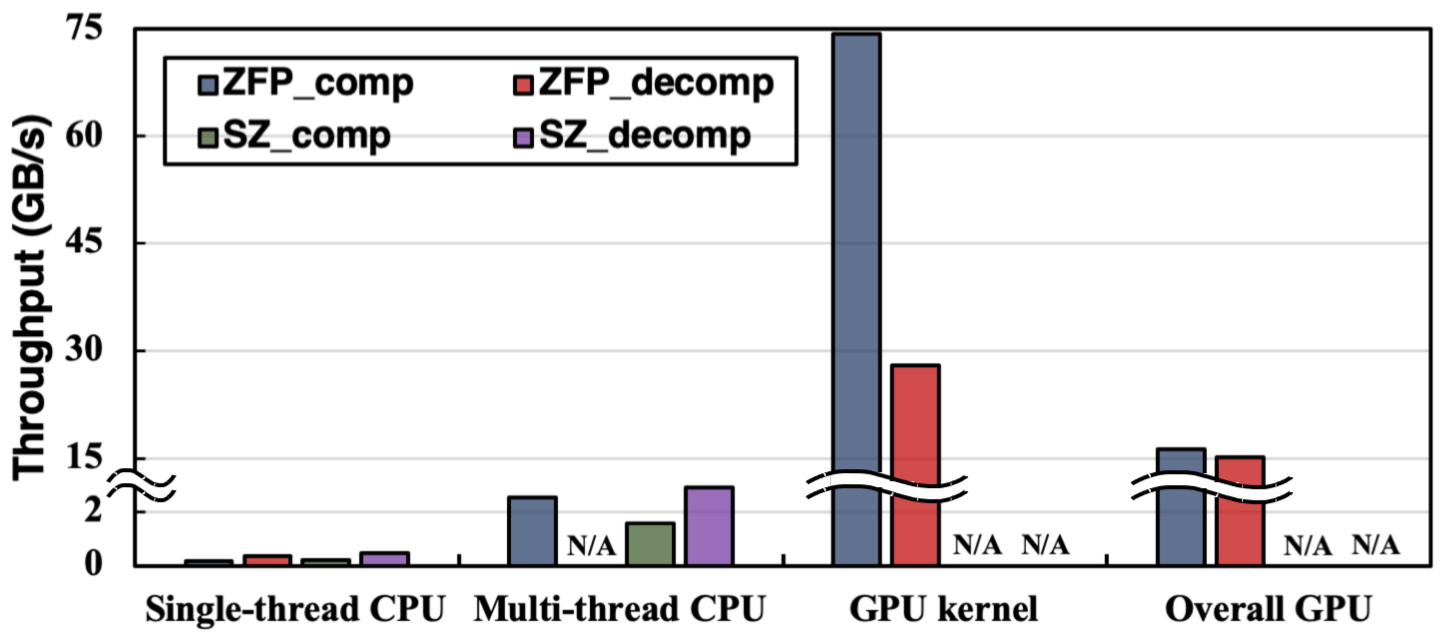}
\centering
\vspace{-4mm}
\caption{Comparison of compression and decompression throughput with SZ and ZFP using Intel Xeon Gold 6148 CPU and Nvidia Tesla V100 GPU on PantaRhei cluster.}
\label{fig:throughput-compare}
\vspace{-6mm}
\end{figure}

Because we already determined the best-fit compression configurations for cuZFP and GPU-SZ on the Nyx dataset in the last section, we will keep using those settings for the overall throughput evaluation. 
Specifically, we set the bitrate of $(4, 4, 4, 2, 2, 2)$ and the absolute error bound of $(0.2, 0.4, 1\text{e+}3, 2\text{e+}5, 2\text{e+}5, 2\text{e+}5)$ for the six Nyx fields with cuZFP and with GPU-SZ, respectively. 
Figure~\ref{fig:throughput-compare} shows the comparison of compression and decompression throughput using SZ and ZFP on CPU and GPU. 
We use the Nvidia Tesla V100 GPU and the 20-core Intel Xeon Gold 6148 CPU in this evaluation.
Note that ZFP does not support the decompression with OpenMP yet~\cite{zfp-doc} (i.e., denoted by ``N/A''). 
We observe that even considering the time of data transfer between CPU and GPU, the GPU-based lossy compression such as cuZFP can still achieve much higher throughput than does the compression on a multi-core CPU.
Moreover, taking into account multiple GPUs on a single node, for instance, six Nvidia Tesla V100 GPUs per Summit node, cuZFP can significantly reduce the compression overhead to $\frac{1}{40}$ of the original multi-core compression overhead (e.g., from more than 10\% to lower than 0.3\%).
Again, the overall compression and decompression throughput can be further improved by using a faster CPU-GPU interconnect or asynchronous GPU-CPU communication.
We will further evaluate our work on the advanced supercomputers such as Summit in the future.

\begin{figure}[]
\includegraphics[width=8.8cm]{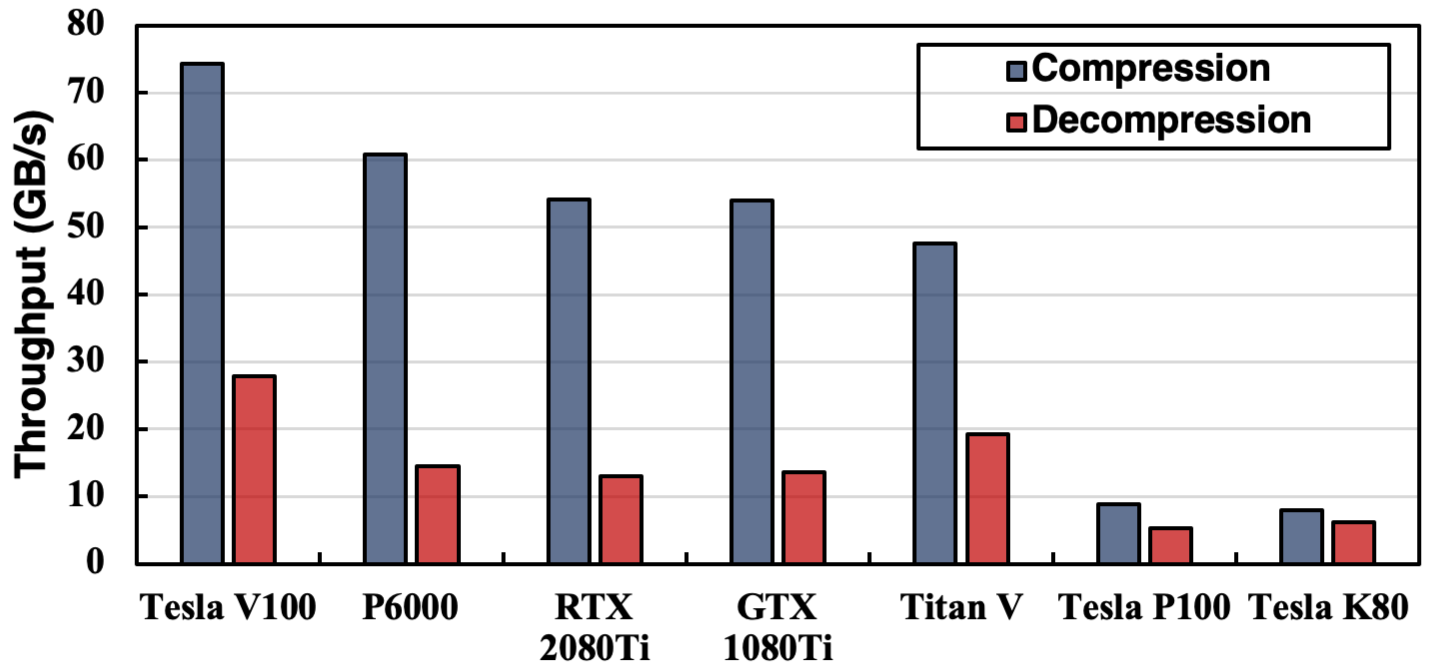}
\centering
\vspace{-6mm}
\caption{Comparison of compression and decompression throughput with cuZFP on different GPUs.}
\label{fig:different-gpus}
\vspace{-6mm}
\end{figure}

We also conducted an experiment to evaluate the compression and decompression kernel throughput on different GPUs. The data transfer time for different GPUs is similar, because all the GPUs are connected to their hosts via 16-lane PCIe 3.0 interconnect. 
Figure~\ref{fig:different-gpus} and Table~\ref{tab:gpus} demonstrate that the kernel throughputs can be increased by using upgraded GPU hardware, such as more shaders, higher peak performance, and higher memory bandwidth. 

\subsection{Configuration Optimization Guideline}

To determine the best-fit compression configurations for different lossy compressors and cosmological datasets, we need to balance three key features: compression quality (postanalysis), compression ratio (storage), and overall compression throughput (performance).
In the above section, we analyzed the relationship between compression quality and ratio based on the tested datasets. In the following discussion, we will explain how they interact with overall throughput.

\begin{figure}[]
\includegraphics[width=0.9\linewidth]{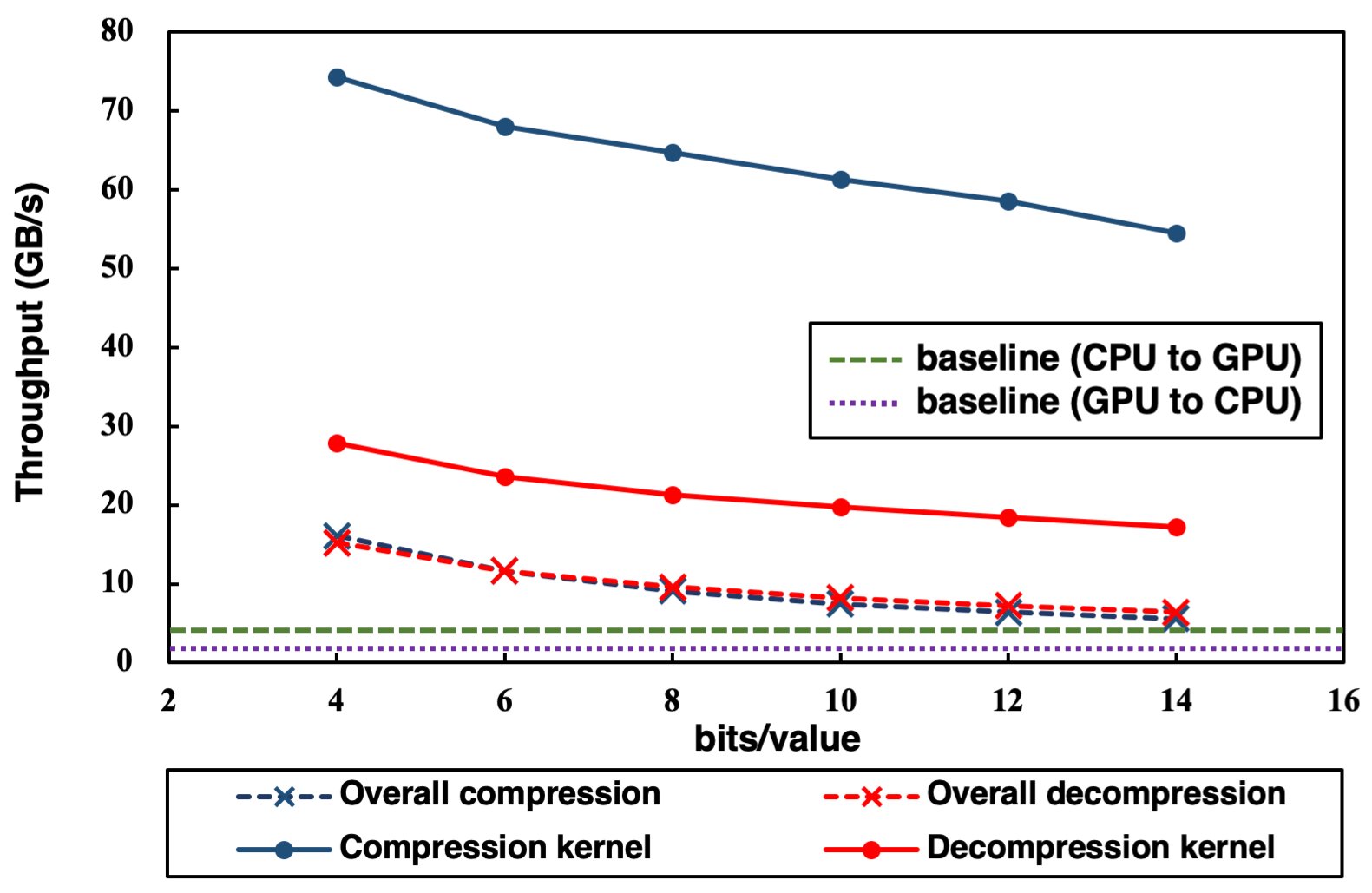}
\centering
\caption{Compression and decompression throughput with cuZFP on Nyx dataset. \small{Dashed lines represent throughput including data transfer between CPU and GPU. Solid lines represent cuZFP kernel throughput. Baseline is data transfer without compression.}}
\label{fig:throughput-bitrate}
\vspace{-6mm}
\end{figure}

Figure~\ref{fig:throughput-bitrate} illustrates the relationship between overall throughput and bitrate with cuZFP on the Nyx dataset. We can observe that both the kernel throughput and overall throughput are decreased by increasing the bitrate. 
Higher bitrate (i.e., lower compression ratio) usually means storing more useful information (as illustrated in Figure~\ref{fig:rate-distortion}), which would lead to more computations during compression and decompression. This phenomenon is also observed in many previous studies~\cite{tao2017significantly, zfp}.
Also, as we discussed above, the data transfer time between CPU and GPU also increases with bitrate.
Therefore, this suggests that after determining a set of compression configurations (e.g., bitrate or error bound) that can generate acceptable reconstructed data considering power spectrum and halo finder, we can choose the one with the highest compression ratio to be the best-fit configuration, which can result in both the highest overall throughput and performance and lowest storage overhead. 

In summary, our optimization guideline consists of (1) using our Foresight framework to benchmark different GPU-based lossy compressors with various configurations based on the target cosmological simulation datasets;
(2) identifying a set of configurations to produce acceptable reconstructed data based on the power spectrum and halo finder analysis using Foresight; 
and (3) choosing the configuration with the highest compression ratio as the best-fit setting. 
By using these guidelines, cosmology researchers and scientists can find the best-fit compression configuration in terms of both performance and storage overhead for using the in situ lossy compression in their extreme-scale cosmological simulation.
For scientific applications in other domains, we can apply our approach to them by simply implementing additional analysis functions and/or scripts to the Foresight's PAT component.
\section{Related Work}
\label{sec:related}

Although lossy compression has been studied for many years to save the storage space and I/O cost for large-scale scientific simulations, such as climate simulation~\cite{baker2014methodology,baker2017toward}, it has been recently investigated for extreme-scale cosmological simulations. 
For example, Tao et al.~\cite{tao2017exploration,tao2017depth} proposed to use the data smoothness in space to improve the SZ lossy compressor for single snapshots of cosmological simulations.
A few studies~\cite{tao2017exploration,tao2017exploration,li2018optimizing} later identified that cosmological simulation data have very low smoothness in space, leading to low compression quality. Thus, Li et al.~\cite{li2018optimizing} proposed an optimized compression model that combines space-based compression and time-based compression for cosmological simulations.
However, unlike the lossy compression studies for climate simulations that have investigated the evaluation metrics suggested by climatologists~\cite{baker2014methodology}, existing work for cosmological simulations has only considered general compression quality metrics, such as compression ratio, rate-distortion, and compression and decompression throughput, all of which cannot guarantee the feasibility of using lossy compression for real-world cosmological simulations, as discussed in Section~\ref{sec:problem}.
Additionally, none of the existing work has studied advanced GPU-based lossy compressors for large-scale scientific simulations, including cosmological simulations. 
Compared with existing work, our paper is the first work on quantitatively evaluating GPU-based lossy compression on real-world extreme-scale cosmological simulations that take into account both general and cosmology-specific evaluation metrics suggested by domain scientists.
\section{Conclusion and Future Work}
\label{sec:conclusion}

In this paper, we conduct a thorough empirical evaluation for two leading GPU-based error-bounded lossy compressors on the real-world extreme-scale cosmological simulation datasets HACC and Nyx. We evaluate different compression configurations and how they affect the general compression quality and post-analysis quality.
For easier evaluation, we implement the GPU-based lossy compressors into our open-source compression benchmark and analysis tool Foresight.
Finally, we provide a general optimization guideline for cosmologists to determine the best-fit configurations for different GPU-based lossy compressors and extreme-scale cosmological simulations.
In the future, we plan to investigate more GPU-based lossy compressors once they are released and more real-world extreme-scale scientific simulations. 
\section*{Acknowledgment}
\scriptsize

This research was supported by the Exasky Exascale Computing Project (ECP), Project Number: 17-SC-20-SC, a collaborative effort of two DOE organizations---the Office of Science and the National Nuclear Security Administration, responsible for the planning and preparation of a capable exascale ecosystem, including software, applications, hardware, advanced system engineering and early testbed platforms, to support the nation's exascale computing imperative. The material was supported by the U.S. Department of Energy, Office of Science, under contract DE-AC02-06CH11357, and supported by the U.S. National Science Foundation under Grants OAC-1948447 and OAC-2034169, and the U.S. National Oceanic and Atmospheric Administration under Grant NA18NWS46200438. We acknowledge the computing resources provided on Darwin, which is operated by the Los Alamos National Laboratory.

\bibliographystyle{IEEEtran}
\bibliography{refs}

\begin{thebibliography}{10}
\providecommand{\url}[1]{#1}
\csname url@samestyle\endcsname
\providecommand{\newblock}{\relax}
\providecommand{\bibinfo}[2]{#2}
\providecommand{\BIBentrySTDinterwordspacing}{\spaceskip=0pt\relax}
\providecommand{\BIBentryALTinterwordstretchfactor}{4}
\providecommand{\BIBentryALTinterwordspacing}{\spaceskip=\fontdimen2\font plus
\BIBentryALTinterwordstretchfactor\fontdimen3\font minus
  \fontdimen4\font\relax}
\providecommand{\BIBforeignlanguage}[2]{{%
\expandafter\ifx\csname l@#1\endcsname\relax
\typeout{** WARNING: IEEEtran.bst: No hyphenation pattern has been}%
\typeout{** loaded for the language `#1'. Using the pattern for}%
\typeout{** the default language instead.}%
\else
\language=\csname l@#1\endcsname
\fi
#2}}
\providecommand{\BIBdecl}{\relax}
\BIBdecl

\bibitem{heitmann2019hacc}
K.~Heitmann, T.~D. Uram, H.~Finkel, N.~Frontiere, S.~Habib, A.~Pope, E.~Rangel,
  J.~Hollowed, D.~Korytov, P.~Larsen, B.~S. Allen, K.~Chard, and I.~Foster,
  ``Hacc cosmological simulations: First data release,'' \emph{arXiv preprint
  arXiv:1904.11966}, 2019.

\bibitem{summit}
{Summit supercomputer}, \url{https://www.olcf.ornl.gov/summit/}.

\bibitem{almgren2013nyx}
A.~S. Almgren, J.~B. Bell, M.~J. Lijewski, Z.~Luki{\'c}, and E.~Van~Andel,
  ``Nyx: A massively parallel amr code for computational cosmology,'' \emph{The
  Astrophysical Journal}, vol. 765, no.~1, p.~39, 2013.

\bibitem{habib2013hacc}
S.~Habib, V.~Morozov, N.~Frontiere, H.~Finkel, A.~Pope, and K.~Heitmann,
  ``Hacc: extreme scaling and performance across diverse architectures,'' in
  \emph{Proceedings of the International Conference on High Performance
  Computing, Networking, Storage and Analysis}.\hskip 1em plus 0.5em minus
  0.4em\relax ACM, 2013, p.~6.

\bibitem{habib2016hacc}
S.~Habib, A.~Pope, H.~Finkel, N.~Frontiere, K.~Heitmann, D.~Daniel, P.~Fasel,
  V.~Morozov, G.~Zagaris, T.~Peterka, V.~Venkatram, L.~Zarija, S.~Saba, and
  W.-k. Liao, ``Hacc: Simulating sky surveys on state-of-the-art supercomputing
  architectures,'' \emph{New Astronomy}, vol.~42, pp. 49--65, 2016.

\bibitem{wan2017comprehensive}
L.~Wan, M.~Wolf, F.~Wang, J.~Y. Choi, G.~Ostrouchov, and S.~Klasky,
  ``Comprehensive measurement and analysis of the user-perceived i/o
  performance in a production leadership-class storage system,'' in \emph{2017
  IEEE 37th International Conference on Distributed Computing Systems
  (ICDCS)}.\hskip 1em plus 0.5em minus 0.4em\relax IEEE, 2017, pp. 1022--1031.

\bibitem{wan2017analysis}
------, ``Analysis and modeling of the end-to-end i/o performance on olcf's
  titan supercomputer,'' in \emph{2017 IEEE 19th International Conference on
  High Performance Computing and Communications; IEEE 15th International
  Conference on Smart City; IEEE 3rd International Conference on Data Science
  and Systems (HPCC/SmartCity/DSS)}.\hskip 1em plus 0.5em minus 0.4em\relax
  IEEE, 2017, pp. 1--9.

\bibitem{cappello2019use}
F.~Cappello, S.~Di, S.~Li, X.~Liang, A.~M. Gok, D.~Tao, C.~H. Yoon, X.-C. Wu,
  Y.~Alexeev, and F.~T. Chong, ``Use cases of lossy compression for
  floating-point data in scientific data sets,'' \emph{The International
  Journal of High Performance Computing Applications}, 2019.

\bibitem{tao2017significantly}
D.~Tao, S.~Di, Z.~Chen, and F.~Cappello, ``Significantly improving lossy
  compression for scientific data sets based on multidimensional prediction and
  error-controlled quantization,'' in \emph{2017 IEEE International Parallel
  and Distributed Processing Symposium}.\hskip 1em plus 0.5em minus 0.4em\relax
  IEEE, 2017, pp. 1129--1139.

\bibitem{di2016fast}
S.~Di and F.~Cappello, ``Fast error-bounded lossy {HPC} data compression with
  {SZ},'' in \emph{2016 IEEE International Parallel and Distributed Processing
  Symposium}.\hskip 1em plus 0.5em minus 0.4em\relax IEEE, 2016, pp. 730--739.

\bibitem{liangerror}
X.~Liang, S.~Di, D.~Tao, S.~Li, S.~Li, H.~Guo, Z.~Chen, and F.~Cappello,
  ``Error-controlled lossy compression optimized for high compression ratios of
  scientific datasets,'' 2018.

\bibitem{zfp}
P.~Lindstrom, ``Fixed-rate compressed floating-point arrays,'' \emph{IEEE
  Transactions on Visualization and Computer Graphics}, vol.~20, no.~12, pp.
  2674--2683, 2014.

\bibitem{lu2018understanding}
T.~Lu, Q.~Liu, X.~He, H.~Luo, E.~Suchyta, J.~Choi, N.~Podhorszki, S.~Klasky,
  M.~Wolf, T.~Liu, and Z.~Qiao, ``Understanding and modeling lossy compression
  schemes on {HPC} scientific data,'' in \emph{2018 IEEE International Parallel
  and Distributed Processing Symposium}.\hskip 1em plus 0.5em minus 0.4em\relax
  IEEE, 2018, pp. 348--357.

\bibitem{luo2019identifying}
H.~Luo, D.~Huang, Q.~Liu, Z.~Qiao, H.~Jiang, J.~Bi, H.~Yuan, M.~Zhou, J.~Wang,
  and Z.~Qin, ``Identifying latent reduced models to precondition lossy
  compression,'' in \emph{2019 IEEE International Parallel and Distributed
  Processing Symposium}.\hskip 1em plus 0.5em minus 0.4em\relax IEEE, 2019.

\bibitem{tao2018optimizing}
D.~Tao, S.~Di, X.~Liang, Z.~Chen, and F.~Cappello, ``Optimizing lossy
  compression rate-distortion from automatic online selection between sz and
  zfp,'' \emph{IEEE Transactions on Parallel and Distributed Systems}, 2019.

\bibitem{eke2001power}
V.~R. Eke, J.~F. Navarro, and M.~Steinmetz, ``The power spectrum dependence of
  dark matter halo concentrations,'' \emph{The Astrophysical Journal}, vol.
  554, no.~1, p. 114, 2001.

\bibitem{hacc-summit}
H.~Salman, ``Marching to exascale: Extreme-scale cosmological simulations with
  hacc on summit,''
  \url{https://www.olcf.ornl.gov/wp-content/uploads/2018/10/habib_2019OLCFUserMeeting.pdf},
  2019.

\bibitem{sz-openmp-report}
{SZ development team}, ``Openmp version of sz and evaluation report,''
  \url{http://tao.cs.ua.edu/paper/SZ-openmp-evaluation-report.pdf}, 2018.

\bibitem{cesm-simulation}
{Community Earth Simulation Model}, \url{http://www.cesm.ucar.edu}.

\bibitem{baker2014methodology}
A.~H. Baker, H.~Xu, J.~M. Dennis, M.~N. Levy, D.~Nychka, S.~A. Mickelson,
  J.~Edwards, M.~Vertenstein, and A.~Wegener, ``A methodology for evaluating
  the impact of data compression on climate simulation data,'' in
  \emph{Proceedings of the 23rd International Symposium on High-Performance
  Parallel and Distributed Computing}.\hskip 1em plus 0.5em minus 0.4em\relax
  ACM, 2014, pp. 203--214.

\bibitem{lindstrom2006fast}
P.~Lindstrom and M.~Isenburg, ``Fast and efficient compression of
  floating-point data,'' \emph{IEEE Transactions on Visualization and Computer
  Graphics}, vol.~12, no.~5, pp. 1245--1250, 2006.

\bibitem{FPC}
M.~Burtscher and P.~Ratanaworabhan, ``Fpc: A high-speed compressor for
  double-precision floating-point data,'' \emph{IEEE Transactions on
  Computers}, vol.~58, no.~1, pp. 18--31, 2008.

\bibitem{son2014data}
S.~W. Son, Z.~Chen, W.~Hendrix, A.~Agrawal, W.-k. Liao, and A.~Choudhary,
  ``Data compression for the exascale computing era-survey,''
  \emph{Supercomputing Frontiers and Innovations}, vol.~1, no.~2, pp. 76--88,
  2014.

\bibitem{chandra2001parallel}
R.~Chandra, L.~Dagum, D.~Kohr, R.~Menon, D.~Maydan, and J.~McDonald,
  \emph{Parallel programming in OpenMP}.\hskip 1em plus 0.5em minus 0.4em\relax
  Morgan kaufmann, 2001.

\bibitem{cuZFP}
{cuZFP}, \url{https://github.com/LLNL/zfp/tree/develop/src/cuda_zfp}.

\bibitem{nyx}
{Nyx}, \url{https://github.com/AMReX-Astro/Nyx}.

\bibitem{liang2018efficient}
X.~Liang, S.~Di, D.~Tao, Z.~Chen, and F.~Cappello, ``An efficient
  transformation scheme for lossy data compression with point-wise relative
  error bound,'' in \emph{2018 IEEE International Conference on Cluster
  Computing}.\hskip 1em plus 0.5em minus 0.4em\relax IEEE, 2018, pp. 179--189.

\bibitem{Davis1985}
M.~Davis, G.~Efstathiou, C.~S. Frenk, and S.~D. White, ``The evolution of
  large-scale structure in a universe dominated by cold dark matter,''
  \emph{The Astrophysical Journal}, vol. 292, pp. 371--394, 1985.

\bibitem{foresight-git}
C.~Biwer, P.~Grosset, S.~Jin, J.~Pulido, and H.~Rakotoarivelo,
  ``Vizaly-foresight: A compression benchmark suite for visualization and
  analysis of simulation data,''
  \url{https://github.com/lanl/VizAly-Foresight}.

\bibitem{da2017characterization}
R.~F. da~Silva, R.~Filgueira, I.~Pietri, M.~Jiang, R.~Sakellariou, and
  E.~Deelman, ``A characterization of workflow management systems for
  extreme-scale applications,'' \emph{Future Generation Computer Systems},
  vol.~75, pp. 228--238, 2017.

\bibitem{yoo2003slurm}
A.~B. Yoo, M.~A. Jette, and M.~Grondona, ``Slurm: Simple linux utility for
  resource management,'' in \emph{Workshop on Job Scheduling Strategies for
  Parallel Processing}.\hskip 1em plus 0.5em minus 0.4em\relax Springer, 2003,
  pp. 44--60.

\bibitem{Woodring:2017}
J.~Woodring, J.~P. Ahrens, J.~Patchett, C.~Tauxe, and D.~H. Rogers,
  ``High-dimensional scientific data exploration via cinema,'' in \emph{2017
  IEEE Workshop on Data Systems for Interactive Analysis}.\hskip 1em plus 0.5em
  minus 0.4em\relax IEEE, 2017, pp. 1--5.

\bibitem{sz-openmp}
{SZ Lossy Compressor}, \url{https://github.com/disheng222/SZ}.

\bibitem{zg3m-8j73-19}
\BIBentryALTinterwordspacing
K.~Heitmann, ``Timestep 499 of small outer rim,'' 2019. [Online]. Available:
  \url{http://dx.doi.org/10.21227/zg3m-8j73}
\BIBentrySTDinterwordspacing

\bibitem{genericio}
{GenericIO}, \url{https://trac.alcf.anl.gov/projects/genericio}.

\bibitem{k8gb-vq78-19}
\BIBentryALTinterwordspacing
Z.~Lukic, ``Nyx cosmological simulation data,'' 2019. [Online]. Available:
  \url{http://dx.doi.org/10.21227/k8gb-vq78}
\BIBentrySTDinterwordspacing

\bibitem{folk1999hdf5}
M.~Folk, A.~Cheng, and K.~Yates, ``Hdf5: A file format and i/o library for high
  performance computing applications,'' in \emph{Proceedings of
  supercomputing}, vol.~99, 1999, pp. 5--33.

\bibitem{darwin}
{Darwin cluster}, \url{https://www.osti.gov/biblio/1441285-darwin-cluster}.

\bibitem{pantarhei}
{PantaRhei cluster}, \url{https://www.dingwentao.com/experimental-system}.

\bibitem{tao2017depth}
D.~Tao, S.~Di, Z.~Chen, and F.~Cappello, ``In-depth exploration of
  single-snapshot lossy compression techniques for n-body simulations,'' in
  \emph{2017 IEEE International Conference on Big Data}.\hskip 1em plus 0.5em
  minus 0.4em\relax IEEE, 2017, pp. 486--493.

\bibitem{li2018optimizing}
S.~Li, S.~Di, X.~Liang, Z.~Chen, and F.~Cappello, ``Optimizing lossy
  compression with adjacent snapshots for n-body simulation data,'' in
  \emph{2018 IEEE International Conference on Big Data}.\hskip 1em plus 0.5em
  minus 0.4em\relax IEEE, 2018, pp. 428--437.

\bibitem{zfp-doc}
P.~Lindstrom, M.~Salasoo, M.~Larsen, and S.~Herbein, ``zfp documentation,''
  \url{https://buildmedia.readthedocs.org/media/pdf/zfp/latest/zfp.pdf}.

\bibitem{baker2017toward}
A.~H. Baker, H.~Xu, D.~M. Hammerling, S.~Li, and J.~P. Clyne, ``Toward a
  multi-method approach: Lossy data compression for climate simulation data,''
  in \emph{International Conference on High Performance Computing}.\hskip 1em
  plus 0.5em minus 0.4em\relax Springer, 2017, pp. 30--42.

\bibitem{tao2017exploration}
D.~Tao, S.~Di, Z.~Chen, and F.~Cappello, ``Exploration of pattern-matching
  techniques for lossy compression on cosmology simulation data sets,'' in
  \emph{International Conference on High Performance Computing}.\hskip 1em plus
  0.5em minus 0.4em\relax Springer, 2017, pp. 43--54.

\end{thebibliography}

\end{document}